\documentclass[12pt]{article}

\usepackage{graphicx,amssymb,amsmath,dsfont,txfonts} 
\usepackage{bbm} 

\usepackage[a4paper]{geometry} 
\def\dfrac#1#2{\frac{\displaystyle\strut #1}{\displaystyle\strut #2}}
\DeclareMathOperator{\tr}{tr}
\def\bra#1{\mathinner{\langle{#1}|}}
\def\ket#1{\mathinner{|{#1}\rangle}}
\def\braket#1{\mathinner{\langle{#1}\rangle}}
\newcommand{\eM}{M}

\begin{document}

\title{\LARGE\bf 
Noisy quantum metrology \\ with the assistance of indefinite causal order}

\author{Fran\c{c}ois {\sc Chapeau-Blondeau}, \\
    Laboratoire Angevin de Recherche en Ing\'enierie des Syst\`emes (LARIS), \\
    Universit\'e d'Angers,
    62 avenue Notre Dame du Lac, 49000 Angers, France.
}

\date{\today}

\maketitle

\parindent=8mm \parskip=0ex

\begin{abstract}
A generic qubit unitary operator affected by depolarizing noise is duplicated and inserted in 
a quantum switch process realizing a superposition of causal orders. The characterization of the 
resulting switched quantum channel is worked out for its action on the joint state of 
the probe-control qubit pair. The switched channel is then specifically investigated for the 
important metrological task of phase estimation on the noisy unitary operator, with the 
performance assessed by the Fisher information, classical or quantum.
A comparison is made with conventional techniques of estimation where the noisy unitary is 
directly probed in a one-stage or two-stage cascade with definite order, or several uses of 
them with two or more qubits.
In the switched channel with indefinite order, specific properties are reported, meaningful for 
estimation and not present with conventional techniques. It is shown that the control qubit, 
although it never directly interacts with the unitary, can nevertheless be measured alone for 
effective estimation, while discarding the probe qubit that interacts with the unitary.
Also, measurement of the control qubit maintains the possibility of efficient estimation 
in difficult conditions where conventional estimation becomes less efficient, as for instance 
with ill-configured input probes, or in blind situations when the axis of the unitary is unknown.
Especially, effective estimation by measuring the control qubit remains possible even when the 
input probe tends to align with the axis of the unitary, or with a fully depolarized input probe, 
while in these conditions conventional estimation becomes inoperative.
Measurement of the probe qubit of the switched channel is also analyzed and shown to add useful 
capabilities for phase estimation. The results contribute to the ongoing identification and 
analysis of the properties and capabilities of switched quantum channels with indefinite order 
for information processing, and uncover new possibilities for quantum estimation and qubit 
metrology.
\end{abstract}

\section{Introduction}

{\let\thefootnote\relax\footnote{{Preprint of a paper published by {\em Physical Review A},
vol.~103, 032615, pp.~1-18 (2021). \\
https://doi.org/10.1103/PhysRevA.103.032615 \hfill 
https://journals.aps.org/pra/abstract/10.1103/PhysRevA.103.032615}}}
Quantum channels can be viewed as building blocks for performing quantum information processing 
by transforming quantum states or signals, much like in classical systems-and-signals theory.
Two quantum channels (1) and (2) can be combined or cascaded, in the order (1)--(2) or
(2)--(1), under the control of a quantum switch signal realized for instance by the two basis 
states of a qubit. Such a control qubit can be placed in an arbitrary superposition of its 
two basis states, and as a result, the two individual channels get cascaded in an arbitrary 
superposition of the two classical orders (1)--(2) or (2)--(1). This realizes a switched quantum 
channel formed by the two individual channels simultaneously cascaded in the two alternative 
orders, or with indefinite causal order. Such switched quantum channels with indefinite causal 
order are specifically quantum devices, grounded on quantum superposition, and with no classical 
analogue. Their principle has been described recently in \cite{Oreshkov12,Chiribella13} and their 
physical implementation is addressed in 
\cite{Chiribella13,Procopio15,Rubino17,Goswami18,Wei19,Guo20}.
For information processing, it is being found that switched quantum channels with indefinite 
causal order can specifically offer useful capabilities, not accessible with channels combined 
in definite causal orders. Such specific capabilities from switched indefinite causal order have
been reported with various channels and information processing tasks, assessed by appropriate 
efficiency metrics. 

For instance, Refs.~\cite{Ebler18,Procopio19,Procopio20,Loizeau20} address a task of quantum 
communication of information, where typically isolated communication channels with limited 
capacity, when inserted in a quantum switch process, give rise to a switched quantum channel 
with enhanced capacity to transmit information.
The task in \cite{Chiribella12,Koudia19} is quantum channel discrimination; 
in \cite{Chiribella12} two channels that when associated in definite order are never perfectly 
distinguishable become so with indefinite order; 
in \cite{Koudia19} for distinguishing whether or not a qubit has been affected by a given 
unitary transformation in the presence of noise, the probability of discrimination error is 
shown improvable by the quantum switch process.
Very recently, switched channels with indefinite order have been extended to quantum metrological
tasks involving parameter estimation from measurements \cite{Frey19,Mukhopadhyay18,Zhao20}.
It has been shown that a one-parameter quantum channel can be identified or estimated more 
efficiently when it is involved in a switch process, for a qudit depolarizing channel in \cite{Frey19}
and a qubit thermalization channel in \cite{Mukhopadhyay18}.
In \cite{Zhao20}, to estimate the product of two average displacements in a continuous-variable 
quantum system, it is shown that the estimation error can be reduced by the switch process.

Quantum switching with indefinite causal order is a phenomenon of recent introduction, and its 
properties and capabilities for information processing are still being inventoried and analyzed.
In the present paper, we will extend the investigation of switched causal orders for quantum 
metrology, applied here to new quantum processes or channels and in different conditions. We will 
address the important task of quantum metrology consisting in parameter estimation on a unitary 
transformation in the presence of noise \cite{Giovannetti06,Shaji07,Giovannetti11}. 
For quantum metrology, the switched processes considered here are different from those of 
\cite{Frey19,Mukhopadhyay18,Zhao20}, and the presence of noise is a significant specificity here.
The Fisher information will be used to assess the performance for estimation, as in
\cite{Frey19,Mukhopadhyay18,Zhao20}; the Fisher information being a fundamental reference metric 
in metrology, often employed for characterization and fixing the best conceivable performance
\cite{Barndorff00,Paris09}.

In this paper, we will first briefly review the principle of the quantum switch of elementary
channels. Then, from an elementary channel formed by a generic qubit unitary operator affected by 
depolarizing noise, we will carry out a complete characterization of the transformation realized 
by the corresponding switched quantum channel, specially by means of the Bloch representation of 
the qubit density operator. We will then concentrate on the important metrological task of 
estimating the phase parameter of the unitary operator, and evaluate the Fisher information to 
assess the performance. We will show the possibility of useful properties for the phase estimation 
with indefinite causal order in the switched unitary channel with noise. Especially, conditions 
will be reported where the switched channel with indefinite order remains efficient for estimation, 
while conventional techniques with definite order become inoperative. The study in this way will 
extend the analysis of switched channels with indefinite causal order to the task of estimation 
on a noisy qubit unitary, and will report new possibilities to contribute to quantum estimation 
and qubit metrology.

\section{Quantum switch of two quantum channels} \label{switch_sec}

\begin{figure}[htb]
\centerline{\includegraphics[width=70mm]{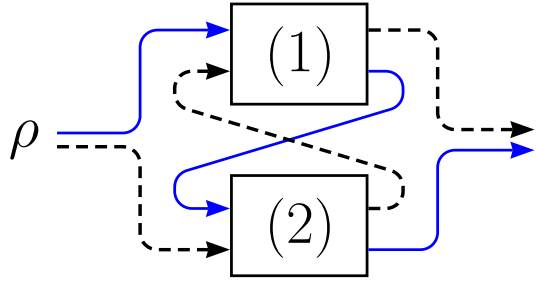}}
\caption[what appears in lof LL p177]
{Two quantum channels (1) and (2) can be cascaded either in the order (1)--(2) (solid path) 
or (2)--(1) (dashed path) according to the state respectively $\ket{0}$ or $\ket{1}$ of a control 
qubit.
}
\label{figSwi1}
\end{figure}

We consider as in \cite{Chiribella13,Ebler18}, acting on quantum systems with Hilbert space 
$\mathcal{H}$, a quantum channel (1) with Kraus operators $\bigl\{\mathsf{K}_k^{(1)} \bigr\}$ and 
a second quantum channel (2) with Kraus operators $\bigl\{\mathsf{K}_j^{(2)} \bigr\}$. The two 
channels are cascaded either in the order (1)--(2) or (2)--(1), under the control of a qubit 
driving a quantum switch process as described for instance in \cite{Chiribella13,Ebler18}. When 
the control qubit is in state $\ket{0}$ channel (1) is traversed first, followed by channel (2); 
and when the control qubit is in state $\ket{1}$ channel (2) is traversed first, followed by 
channel (1), as depicted in Fig.~\ref{figSwi1}.

The resulting ``switched'' quantum channel is described \cite{Chiribella13,Ebler18} by the Kraus 
operators 
\begin{equation}
\mathsf{K}_{jk} = \mathsf{K}_j^{(2)} \mathsf{K}_k^{(1)} \otimes \ket{0}\bra{0}_c + 
                  \mathsf{K}_k^{(1)} \mathsf{K}_j^{(2)} \otimes \ket{1}\bra{1}_c \;.
\label{Wjk}
\end{equation}
When acting on a quantum state of $\mathcal{H}$ with density operator $\rho$ along with a
control qubit in state $\rho_c$, the switched quantum channel implements the bipartite quantum 
operation $\mathcal{S}$ defined \cite{Chiribella13,Ebler18} by the superoperator
\begin{equation}
\mathcal{S}(\rho \otimes \rho_c) = \sum_{j,k} \mathsf{K}_{jk} (\rho \otimes \rho_c) 
\mathsf{K}_{jk}^\dagger \;.
\label{Sgen1}
\end{equation}
The quantum operation realized in Eq.~(\ref{Sgen1}) can be further developed as
\begin{eqnarray}
\nonumber
\mathcal{S}(\rho \otimes \rho_c) &=& 
\mathcal{S}_{00}(\rho) \otimes \braket{0|\rho_c|0}\ket{0}\bra{0}_c +
\mathcal{S}_{01}(\rho) \otimes \braket{0|\rho_c|1}\ket{0}\bra{1}_c \\
\label{Sgen2} 
\mbox{} &+& \mathcal{S}_{01}^\dagger(\rho) \otimes \braket{1|\rho_c|0}\ket{1}\bra{0}_c +
\mathcal{S}_{11}(\rho) \otimes \braket{1|\rho_c|1}\ket{1}\bra{1}_c \;,
\end{eqnarray}
with the superoperators
\begin{eqnarray}
\label{S00}
\mathcal{S}_{00}(\rho) &=& \sum_{j,k} \mathsf{K}_j^{(2)} \mathsf{K}_k^{(1)} \rho 
\mathsf{K}_k^{(1)\dagger} \mathsf{K}_j^{(2)\dagger} \;,\\
\label{S01}
\mathcal{S}_{01}(\rho) &=& \sum_{j,k} \mathsf{K}_j^{(2)} \mathsf{K}_k^{(1)} \rho 
\mathsf{K}_j^{(2)\dagger} \mathsf{K}_k^{(1)\dagger} \;,\\
\label{S11}
\mathcal{S}_{11}(\rho) &=& \sum_{j,k} \mathsf{K}_k^{(1)} \mathsf{K}_j^{(2)} \rho 
\mathsf{K}_j^{(2)\dagger} \mathsf{K}_k^{(1)\dagger} \;.
\end{eqnarray}

The superoperator $\mathcal{S}_{00}(\rho)$ alone describes the quantum operation realized by the 
standard cascade with the definite causal order (1)--(2), and similarly with 
$\mathcal{S}_{11}(\rho)$ for the cascade (2)--(1). By contrast, the superoperator 
$\mathcal{S}_{01}(\rho)$ is a coupling term specific to the quantum switch process. In the joint 
state $\mathcal{S}(\rho \otimes \rho_c)$ of Eq.~(\ref{Sgen2}), if the control qubit were 
discarded (unobserved) and traced out, the resulting quantum operation on $\rho$ would represent 
a classical probabilistic (convex) combination of the two definite causal orders 
$\mathcal{S}_{00}(\rho)$ and $\mathcal{S}_{11}(\rho)$. By contrast, if the control qubit is 
treated coherently with $\rho$ it can give rise to specific, specifically quantum, behaviors 
from the switched quantum channel, as we shall see.

An interesting and specifically quantum feature is that the control qubit can be placed in the 
superposed state $\ket{\psi_c}=\sqrt{p_c}\ket{0}+\sqrt{1-p_c}\ket{1}$, with $p_c \in [0, 1]$.
This produces in Eqs.~(\ref{Sgen1})--(\ref{Sgen2}) a switched quantum channel representing a 
quantum superposition of the two initial channels (1) and (2) simultaneously cascaded in the two 
alternative orders, or with indefinite causal order. With $\rho_c=\ket{\psi_c}\bra{\psi_c}$, the 
quantum operation resulting in Eq.~(\ref{Sgen2}) takes the form
\begin{eqnarray}
\nonumber
\mathcal{S}(\rho \otimes \rho_c) &=& p_c
\mathcal{S}_{00}(\rho) \otimes \ket{0}\bra{0}_c + (1-p_c)
\mathcal{S}_{11}(\rho) \otimes \ket{1}\bra{1}_c \\
\label{Sgenc}
&+& \sqrt{(1-p_c)p_c} \bigl[ \mathcal{S}_{01}(\rho) \otimes \ket{0}\bra{1}_c + 
\mathcal{S}_{01}^\dagger(\rho) \otimes \ket{1}\bra{0}_c \bigr] \;.
\end{eqnarray}

We will consider the situation where the quantum channels (1) and (2) are qubit channels, under a 
form which is often encountered in quantum metrology, and consisting in a unitary operator 
$\mathsf{U}_\xi$ affected by a quantum noise $\mathcal{N}(\cdot)$.

\section{A unitary qubit channel with noise}

For qubits with two-dimensional Hilbert space $\mathcal{H}_2$, the density operator is 
represented in Bloch representation \cite{Nielsen00} under the form
\begin{equation}
\rho =\frac{1}{2}\bigl( \mathrm{I}_2 + \vec{r}\cdot \vec{\sigma} \bigr) \;,
\label{roBloch}
\end{equation}
where $\mathrm{I}_2$ is the identity operator on $\mathcal{H}_2$, and $\vec{\sigma}$ a formal 
vector assembling the three (traceless Hermitian unitary) Pauli operators 
$[\sigma_x, \sigma_y, \sigma_z]=\vec{\sigma}$. The Bloch vector $\vec{r} \in \mathbbm{R}^3$ 
characterizing the density operator has norm $\| \vec{r}\, \|=1$ for a pure state, and 
$\| \vec{r}\, \|<1$ for a mixed state.

A qubit unitary operator $\mathsf{U}_\xi$ is introduced with the general parameterization
\cite{Nielsen00} 
\begin{equation}
\mathsf{U}_\xi=\exp\Bigl(-i\dfrac{\xi}{2} \vec{n} \cdot \vec{\sigma} \Bigr) =
\cos\Bigl(\frac{\xi}{2} \Bigr)\, \mathrm{I}_2 -i \sin\Bigl(\frac{\xi}{2} \Bigr)\,
\vec{n} \cdot \vec{\sigma} \;,
\label{Uxi}
\end{equation}
where $\vec{n}=[n_x, n_y, n_z]^\top$ is a unit vector of $\mathbbm{R}^3$, and $\xi$ a phase 
angle in $[0, 2\pi )$.

From a qubit state $\rho$ in Bloch representation as in Eq.~(\ref{roBloch}), the unitary
$\mathsf{U}_\xi$ produces the transformed state
\begin{equation}
\mathsf{U}_\xi \rho \mathsf{U}_\xi^\dagger = 
\frac{1}{2}\bigl( \mathrm{I}_2 + U_\xi\vec{r}\cdot \vec{\sigma} \bigr) \;,
\label{Uxi_ro}
\end{equation}
which amounts to the transformation $U_\xi\vec{r}$ of the Bloch vector $\vec{r}$ in $\mathbbm{R}^3$
experiencing a rotation around the axis $\vec{n}$ by the angle $\xi$ via the $3\times 3$ real 
matrix\footnote{We use the notation $\mathsf{U}_\xi$ in upright font for the unitary 
operator acting in the complex Hilbert space $\mathcal{H}_2$ of the qubit; while we use the 
notation $U_\xi$ in italic font for the real matrix expressing the action of the unitary 
operator in the Bloch representation of qubit states in $\mathbbm{R}^3$.}
\begin{equation}
U_\xi =
\begin{bmatrix}
\cos(\xi)+[1-\cos(\xi)]n_x^2 & [1-\cos(\xi)]n_x n_y -\sin(\xi)n_z & 
[1-\cos(\xi)]n_x n_z +\sin(\xi)n_y \\
[1-\cos(\xi)]n_x n_y +\sin(\xi)n_z & \cos(\xi)+[1-\cos(\xi)]n_y^2 & 
[1-\cos(\xi)]n_y n_z -\sin(\xi)n_x 
{\vrule height 6mm depth 0mm width 0pt} \\
[1-\cos(\xi)]n_x n_z -\sin(\xi)n_y & [1-\cos(\xi)]n_y n_z +\sin(\xi)n_x & 
\cos(\xi)+[1-\cos(\xi)]n_z^2
{\vrule height 6mm depth 0mm width 0pt}
\end{bmatrix}.
\label{matUxi}
\end{equation}

The qubit noise $\mathcal{N}(\cdot)$ is introduced under the form of a depolarizing noise 
\cite{Nielsen00} implementing the quantum operation with Kraus representation
\begin{equation}
\mathcal{N}(\rho)= (1-p)\rho + \dfrac{p}{3} \bigl(\sigma_x \rho \sigma_x^\dagger 
+ \sigma_y \rho \sigma_y^\dagger + \sigma_z \rho \sigma_z^\dagger \bigr) \;.
\label{depol1}
\end{equation}
The effect of the noise in Eq.~(\ref{depol1}) is to leave the qubit state $\rho$ unchanged 
with the probability $1-p$ or to apply any one of the three Pauli operators with equal 
probability $p/3$. Alternatively, the effect of the depolarizing noise can be described, for the 
Bloch vector $\vec{r}$ characterizing a qubit state in Eq.~(\ref{roBloch}), as the isotropic 
compression $\vec{r} \mapsto \alpha \vec{r}$ with the compression factor $\alpha =1-4p/3$.
Equivalently, Eq.~(\ref{depol1}) is also
\begin{equation}
\mathcal{N}(\rho)= \alpha\rho + (1-\alpha) \frac{\mathrm{I}_2}{2} \;,
\label{depol2}
\end{equation}
indicating that with the probability $1-\alpha$, the noise replaces the quantum state $\rho$ by
the maximally mixed state $\mathrm{I}_2/2$; at the maximum compression when $\alpha =0$ the 
quantum state is forced to $\mathrm{I}_2/2$ with probability $1$ and the qubit gets completely 
depolarized. The depolarizing noise is an important noise model often considered in quantum 
information \cite{Nielsen00}. It has no invariant subspace, and in this respect it represents in 
some sense a worse-case noise and as such a conservative reference. Here, in addition, its 
isotropic character will ease the theoretical derivations. However, this choice for the type of 
noise is not critical for the main properties to be reported here.

\begin{figure}[htb]
\centerline{\includegraphics[width=70mm]{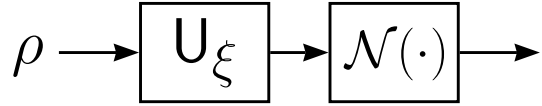}}
\caption[what appears in lof LL p177]
{A qubit channel formed by the unitary transformation $\mathsf{U}_\xi$ of Eq.~(\ref{Uxi}) and 
the depolarizing noise $\mathcal{N}(\cdot)$ of Eqs.~(\ref{depol1})--(\ref{depol2}). As a whole 
this channel is an instance of channel (1) or (2) considered in Fig.~\ref{figSwi1}.
}
\label{figUxiN}
\end{figure}

The quantum channel like (1) or (2) of Section~\ref{switch_sec} is formed by cascading the 
unitary transformation $\mathsf{U}_\xi$ of Eq.~(\ref{Uxi}) and the depolarizing noise 
$\mathcal{N}(\cdot)$ of Eqs.~(\ref{depol1})--(\ref{depol2}), as depicted in Fig.~\ref{figUxiN}.

For the quantum channel of Fig.~\ref{figUxiN}, four Kraus operators like 
$\mathsf{K}_k^{(1)}$ or $\mathsf{K}_j^{(2)}$ of Section~\ref{switch_sec} result as
$\bigl\{\mathsf{K}_0=\sqrt{1-p} \mathsf{U}_\xi, 
\mathsf{K}_1=\sqrt{p/3} \sigma_x\mathsf{U}_\xi,
\mathsf{K}_2=\sqrt{p/3} \sigma_y\mathsf{U}_\xi,
\mathsf{K}_3=\sqrt{p/3} \sigma_z\mathsf{U}_\xi \bigr\}$.  
Equivalently, for a qubit state $\rho$ in Bloch representation as in Eq.~(\ref{roBloch}), 
the cascade of $\mathsf{U}_\xi$ then $\mathcal{N}(\cdot)$ produces the transformed state
\begin{equation}
\mathcal{N}(\mathsf{U}_\xi \rho \mathsf{U}_\xi^\dagger) = 
\frac{1}{2}\bigl( \mathrm{I}_2 + \alpha U_\xi\vec{r}\cdot \vec{\sigma} \bigr) \;.
\label{NUxi_ro}
\end{equation}

We note that, although the four Kraus operators $\mathsf{K}_j$ above generally do not commute 
between them, as a whole the isotropic depolarizing noise $\mathcal{N}(\cdot)$ here commutes with 
the unitary $\mathsf{U}_\xi$ in Fig.~\ref{figUxiN}, so that 
$\mathcal{N}(\mathsf{U}_\xi \rho \mathsf{U}_\xi^\dagger)$ in Eq.~(\ref{NUxi_ro}) coincides with
$\mathsf{U}_\xi \mathcal{N}(\rho) \mathsf{U}_\xi^\dagger$. The noise action $\mathcal{N}(\cdot)$ 
is placed after $\mathsf{U}_\xi$ in Fig.~\ref{figUxiN}, but it could as well take place before 
$\mathsf{U}_\xi$, or even part before and part after $\mathsf{U}_\xi$ and equivalently lumped 
into a single action as in Fig.~\ref{figUxiN}.

\section{Quantum switch of two noisy unitaries} \label{qbswitch_sec}

Two such identical qubit channels formed by $\mathsf{U}_\xi$ and $\mathcal{N}(\cdot)$ as in 
Fig.~\ref{figUxiN} are associated as in Fig.~\ref{figSwi1} through the quantum switch process of 
Section~\ref{switch_sec}, with however two independent noise sources according to 
Eqs.~(\ref{depol1})--(\ref{depol2}) at a same noise level $p$ or $\alpha$. For two identical 
channels (1) and (2), one has $\mathcal{S}_{00}(\rho)=\mathcal{S}_{11}(\rho)$ in 
Eqs.~(\ref{S00}), (\ref{S11}), and also $\mathcal{S}_{01}^\dagger(\rho)=\mathcal{S}_{01}(\rho)$
in Eq.~(\ref{S01}). On the probe qubit in state $\rho$ and control qubit in state
$\rho_c=\ket{\psi_c}\bra{\psi_c}$, the switched quantum channel therefore realizes the two-qubit 
quantum operation from Eq.~(\ref{Sgenc}) reading
\begin{eqnarray}
\nonumber
\mathcal{S}(\rho \otimes \rho_c) &=& \mathcal{S}_{00}(\rho) \otimes \bigl[ 
p_c \ket{0}\bra{0}_c + (1-p_c) \ket{1}\bra{1}_c \bigr]\\
\label{Sgenqb}
&+& \mathcal{S}_{01}(\rho) \otimes \sqrt{(1-p_c)p_c} \bigl( \ket{0}\bra{1}_c + 
\ket{1}\bra{0}_c \bigr) \;.
\end{eqnarray}

Furthermore, as already mentioned, from Eq.~(\ref{S00}) it can be verified that 
$\mathcal{S}_{00}(\rho)$ (and $\mathcal{S}_{11}(\rho)$ similarly) is simply the quantum operation 
realized on $\rho$ by directly traversing in a standard cascade with definite order the two 
channels (1) then (2), which in Bloch representation via Eq.~(\ref{NUxi_ro}) amounts to
\begin{equation}
\mathcal{S}_{00}(\rho)= 
\frac{1}{2}\bigl( \mathrm{I}_2 + \alpha^2 U_\xi^2 \vec{r}\cdot \vec{\sigma} \bigr) \;.
\label{NUNU_ro}
\end{equation}

For the superoperator $\mathcal{S}_{01}(\rho)$ of Eq.~(\ref{S01}) one has now
\begin{eqnarray}
\nonumber
\mathcal{S}_{01}(\rho) &=& (1-p)^2 \mathcal{W}_0 + (1-p)\dfrac{p}{3} 
\sum_{\ell=x,y,z} \bigl(\mathcal{W}_\ell + \mathcal{W}_\ell^\dagger \bigr) 
+ \Bigl(\dfrac{p}{3} \Bigr)^2 \Bigl[ \\
\label{S01_b}
&&
\bigl(\mathcal{W}_{xy} + \mathcal{W}_{xy}^\dagger \bigr) +
\bigl(\mathcal{W}_{yz} + \mathcal{W}_{yz}^\dagger \bigr) +
\bigl(\mathcal{W}_{zx} + \mathcal{W}_{zx}^\dagger \bigr) +
\mathcal{W}_{xx}+\mathcal{W}_{yy}+\mathcal{W}_{zz} \Bigr] \;,
\end{eqnarray}
with the superoperators
\begin{eqnarray}
\label{W0}
\mathcal{W}_0(\rho ) &=& \mathsf{U}_\xi^2 \rho \mathsf{U}_\xi^{\dagger 2} \;, \\
\label{Wl}
\mathcal{W}_\ell(\rho ) &=& \sigma_\ell \mathsf{U}_\xi \mathsf{U}_\xi \rho
\mathsf{U}_\xi^\dagger \sigma_\ell^\dagger \mathsf{U}_\xi^\dagger \;, \\
\label{Wll}
\mathcal{W}_{\ell \ell'}(\rho ) &=& \sigma_\ell \mathsf{U}_\xi \sigma_{\ell'} \mathsf{U}_\xi \rho
\mathsf{U}_\xi^\dagger \sigma_\ell^\dagger \mathsf{U}_\xi^\dagger \sigma_{\ell'}^\dagger \;,
\end{eqnarray}
verifying that $\mathcal{S}_{01}(\rho) =\mathcal{S}_{01}^\dagger(\rho) $.
In Eq.~(\ref{Wl}) and comparable equations, $\sigma_\ell$ designates generically one of the
three Pauli operators $\sigma_x$, $\sigma_y$ or $\sigma_z$ according to the value of $\ell$.
What we want to do next, is to characterize the action of the superoperator 
$\mathcal{S}_{01}(\rho)$ of Eq.~(\ref{S01_b}) by means of the Bloch representation, as in 
Eq.~(\ref{NUNU_ro}). This can be carried out in two steps, by expressing 
$\mathcal{S}_{01}(\mathrm{I}_2)$ and $\mathcal{S}_{01}(\vec{r}\cdot \vec{\sigma} )$.
These derivations are developed in Appendix~A.

From Appendix~A, its Eq.~(\ref{S01_I2_A}), we obtain 
\begin{equation}
\mathcal{S}_{01}(\mathrm{I}_2) =
\biggl[ \dfrac{4}{3}p\Bigl(1-\dfrac{4}{3}p \Bigr)\cos(\xi)
+1-\dfrac{4}{3}p\Bigl(1-\dfrac{p}{3} \Bigr) \biggr] \mathrm{I}_2 \;,
\label{S01_I2}
\end{equation}
and from its Eq.~(\ref{S01_sr2_A}),
\begin{equation}
\mathcal{S}_{01}(\vec{r}\cdot \vec{\sigma}) = \Bigl[
(1-p)^2 U_\xi + (1-p)\dfrac{p}{3} 2 \bigl[I_3+L_1(U_\xi) \bigr]+
\Bigl(\dfrac{p}{3} \Bigr)^2 \bigl[ 2L_2(U_\xi)-2I_3 +L_3(U_\xi) \bigr]
\Bigr] U_\xi \vec{r}\cdot \vec{\sigma} \;.
\label{S01_sr2}
\end{equation}

This completes the characterization of the joint two-qubit state $\mathcal{S}(\rho \otimes \rho_c)$ 
of Eq.~(\ref{Sgenqb}) produced by the switched quantum channel. It follows in particular that, 
when there is no noise, at $p=0$ in Eq.~(\ref{depol1}), the characterization leads to a joint 
state reducing to 
$\mathcal{S}(\rho \otimes \rho_c)=\bigl(\mathsf{U}_\xi^2 \rho \mathsf{U}_\xi^{\dagger 2} \bigr) 
\otimes \rho_c$, indicating that the two qubits evolve separately. The probe qubit in state $\rho$ 
experiences the standard unitary cascade $\mathsf{U}_\xi\mathsf{U}_\xi$, while the control qubit in 
state $\rho_c$ remains unaffected. 
This situation can be understood because with no noise the two channels that are switched are 
two strictly identical unitaries $\mathsf{U}_\xi$, so that the two switched orders 
$\mathsf{U}_\xi\mathsf{U}_\xi$ and $\mathsf{U}_\xi\mathsf{U}_\xi$ in 
Fig.~\ref{figSwi1} are identical and indistinguishable. The resulting switched channel is 
indistinguishable from a standard cascade of two unitaries $\mathsf{U}_\xi$. There is no 
superposition of two distinguishable causal orders, but only a standard cascade with definite 
order. By contrast, in the presence of noise, at $p \not =0$ in Eq.~(\ref{depol1}), the joint state
$\mathcal{S}(\rho \otimes \rho_c)$ of Eq.~(\ref{Sgenqb}) is an entangled state, expressing a 
coupling evolution of the probe-control qubit pair. The two channels according to 
Fig.~\ref{figUxiN} engaged in the switch process of Fig.~\ref{figSwi1}, do not reduce to a standard 
cascade of two indistinguishable channels. 
Indistinguishability of the two switched channels in Fig.~\ref{figSwi1} 
can be attributed to their Kraus operators which do not commute, as in \cite{Ebler18}.
The nonunitary noise process $\mathcal{N}(\cdot)$ shown in Fig.~\ref{figUxiN}, which occurs in two 
independent realizations, involves in Eq.~(\ref{depol1}) Kraus operators which do not commute,
so that the Kraus operators $\mathsf{K}_k^{(1)}$ or $\mathsf{K}_j^{(2)}$ of the two switched 
channels of Fig.~\ref{figUxiN} also do not commute. This induces in Fig.~\ref{figSwi1} a 
superposition of two distinguishable causal orders, and an entangling interaction of the two qubits,
mediated via a nontrivial coupling term $\mathcal{S}_{01}(\rho)$ in the joint state 
$\mathcal{S}(\rho \otimes \rho_c)$ of Eq.~(\ref{Sgenqb}).

We will now examine the exploitation of the joint state $\mathcal{S}(\rho \otimes \rho_c)$ of 
Eq.~(\ref{Sgenqb}) characterizing the switched channel, to serve in a task of parameter 
estimation on the unitary $\mathsf{U}_\xi$.

\section{Measurement} \label{measur_sec}

The probe qubit prepared in state $\rho$ and the control qubit prepared in state $\rho_c$ get 
entangled by the action of the switched quantum channel, and these two qubits together terminate
in the joint state $\mathcal{S}(\rho \otimes \rho_c)$ of Eq.~(\ref{Sgenqb}). To extract 
information from the switched channel, a useful strategy, also adopted for instance in 
\cite{Ebler18}, is to measure the control qubit in the Fourier basis 
$\bigl\{\ket{+}, \ket{-}\bigr\}$ of $\mathcal{H}_2$. The measurement can be described by the two 
measurement operators
$\bigl\{ \mathrm{I}_2 \otimes \ket{+}\bra{+}, \mathrm{I}_2 \otimes \ket{-}\bra{-} \bigr\}$
acting in the Hilbert space $\mathcal{H}_2 \otimes \mathcal{H}_2$ of the probe-control qubit 
pair with state $\mathcal{S}(\rho \otimes \rho_c)$. The measurement randomly projects the control 
qubit either in state $\ket{+}$ or $\ket{-}$, and it leaves the probe qubit in the unnormalized 
conditional state
\begin{equation}
\rho_\pm = {}_c\langle \pm | \mathcal{S}(\rho \otimes \rho_c) | \pm \rangle_c
=\dfrac{1}{2}\mathcal{S}_{00}(\rho) \pm \sqrt{(1-p_c)p_c} \,\mathcal{S}_{01}(\rho) \;,
\label{-S+}
\end{equation}
the products involving $\ket{\pm}_c$ being defined on the control qubit.
The probabilities $P^{\rm con}_\pm$ of the two measurement outcomes are provided by the trace
$P^{\rm con}_\pm =\tr(\rho_\pm )$.
When the probe qubit is prepared in the state $\rho$ of Eq.~(\ref{roBloch}), one has
\begin{equation}
\tr(\rho_\pm )= \dfrac{1}{2} \tr\biggl[
\dfrac{1}{2}\mathcal{S}_{00}(\mathrm{I}_2) \pm \sqrt{(1-p_c)p_c} \,\mathcal{S}_{01}(\mathrm{I}_2)
\biggr] \;,
\label{tr_ro}
\end{equation}
since according to Eqs.~(\ref{NUNU_ro}) and (\ref{S01_sr2}) the terms 
$\mathcal{S}_{00}(\vec{r}\cdot \vec{\sigma})=\alpha^2 U_\xi^2 \vec{r}\cdot \vec{\sigma}$ and
$\mathcal{S}_{01}(\vec{r}\cdot \vec{\sigma})$ are both linear combinations of the three Pauli
operators $\vec{\sigma} $ and are therefore with zero trace. Via Eq.~(\ref{NUNU_ro}) 
giving $\mathcal{S}_{00}(\mathrm{I}_2)=\mathrm{I}_2$ and Eq.~(\ref{S01_I2}) for
$\mathcal{S}_{01}(\mathrm{I}_2)$, one then obtains for the control qubit the measurement
probabilities
\begin{equation}
P^{\rm con}_\pm =\dfrac{1}{2} \pm \sqrt{(1-p_c)p_c}
\biggl[ \dfrac{4}{3}p\Bigl(1-\dfrac{4}{3}p \Bigr)\cos(\xi)
+1-\dfrac{4}{3}p\Bigl(1-\dfrac{p}{3} \Bigr) \biggr] \;,
\label{Pc+}
\end{equation}
which are conveniently rewritten as a function of the compression factor $\alpha$ of the 
depolarizing noise of Eq.~(\ref{depol2}) as
\begin{equation}
P^{\rm con}_\pm =\tr(\rho_\pm )=\dfrac{1}{2} \pm \sqrt{(1-p_c)p_c} Q_\xi(\alpha) \;,
\label{Pc+a}
\end{equation}
with the factor 
\begin{equation}
Q_\xi(\alpha) = (1-\alpha)\alpha\cos(\xi)+\dfrac{1}{4}(1+\alpha)^2 \;.
\label{Qa1}
\end{equation}

In addition, after the measurement of the control qubit, the probe qubit terminates in the
(normalized conditional) state
\begin{equation}
\rho_\pm^{\rm post}=\dfrac{1}{P^{\rm con}_\pm} \rho_\pm =
\frac{1}{2}\bigl( \mathrm{I}_2 + \vec{r}_\pm^{\rm \,post} \cdot \vec{\sigma} \bigr) \;,
\label{ro_post}
\end{equation}
characterized by the post-measurement Bloch vector 
\begin{equation}
\vec{r}_\pm^{\rm \,post} =\dfrac{1}{P^{\rm con}_\pm} \biggl[
\frac{1}{2} \alpha^2 U_\xi^2 
\pm \sqrt{(1-p_c)p_c} \,S_{01} \biggr] \vec{r} \;,
\label{r_post}
\end{equation}
with $S_{01}$ the $3\times 3$ real matrix, quadratic function of $U_\xi$, defined from the 
superoperator $\mathcal{S}_{01}(\vec{r} \cdot \vec{\sigma})=S_{01} \vec{r} \cdot \vec{\sigma}$ 
via Eq.~(\ref{S01_sr2}).

For the control at $p_c=0$ or $1$ there is no superposition of switched orders in 
Fig.~\ref{figSwi1}, but a standard cascade with definite order of two copies of the noisy unitary 
channel of Fig.~\ref{figUxiN} and Eq.~(\ref{NUxi_ro}), yielding $P^{\rm con}_\pm =1/2$ in 
Eq.~(\ref{Pc+a}) and $\vec{r}_\pm^{\rm \,post} =\alpha^2 U_\xi^2 \vec{r}$ in Eq.~(\ref{r_post}). By 
contrast, for any $p_c \in (0, 1)$, some superposition is present in the control signal 
$\ket{\psi_c}$ and therefrom in the two causal orders, inducing in Eq.~(\ref{Pc+a}) a dependence of 
$P^{\rm con}_\pm$ via $Q_\xi(\alpha) $ with the phase $\xi$.

This is an important property that the probabilities $P^{\rm con}_\pm$ of Eq.~(\ref{Pc+a}) upon 
measuring the control qubit, are in general dependent on the phase $\xi$. By measuring the control 
qubit, information can therefore be obtained on the phase $\xi$ and can serve for an estimation of 
$\xi$. It is the probe qubit that directly interacts with the unitary $\mathsf{U}_\xi$ characterized 
by the phase $\xi$, and not the control qubit. However, in the switch process the type of coupling 
between these two qubits in the joint state $\mathcal{S}(\rho \otimes \rho_c)$ of Eq.~(\ref{Sgenqb}), 
causes a transfer of information from the probe to the control qubit concerning the phase $\xi$.

Another important property observed with Eq.~(\ref{Pc+a}) is that the measurement probabilities 
$P^{\rm con}_\pm$ are independent of the Bloch vector $\vec{r}$ characterizing the input probe qubit.
Accordingly, $P^{\rm con}_\pm$ are able to sense the phase $\xi$ in the same way whatever the 
configuration $\vec{r}$ of the input probe, even with a completely depolarized probe with 
$\vec{r}=\vec{0}$.

In a comparable way, the measurement probabilities $P^{\rm con}_\pm$ of Eq.~(\ref{Pc+a}) are 
unaffected by the orientation $\vec{n}$ of the rotation implemented by $\mathsf{U}_\xi$. In this 
respect, $P^{\rm con}_\pm$ can be exploited to estimate the rotation angle $\xi$ equally, even with 
an unknown or an ill-positioned axis $\vec{n}$ relative to the probe $\vec{r}$. This would not be 
the case in a conventional (with no superposition of causal orders) approach of measuring the probe 
qubit to estimate $\xi$, where efficient estimation of $\xi$ would require to know the axis $\vec{n}$ 
and to adjust the estimation conditions (especially the probe $\vec{r}\not =\vec{0}$\,) to this 
$\vec{n}$, as we shall see more precisely below. 

We now concentrate on the task of estimating the phase $\xi$ of the unitary $\mathsf{U}_\xi$.
Phase estimation is an important task of quantum metrology, useful for instance for
interferometry, magnetometry, atomic clocks, frequency standards, and many other high-precision
high-sensitivity physical measurements 
\cite{Giovannetti06,Giovannetti11,DAriano98,vanDam07,Chapeau15,Degen17}. 
A useful tool for assessing and comparing the efficiency of different estimation strategies for 
the phase $\xi$ is provided by the Fisher information, which we now address.

\section{Performance assessment by the Fisher information} \label{Fisher_sec}

Statistical estimation theory \cite{Cover91,Kay93} stipulates that, from data dependent upon a 
parameter $\xi$, any conceivable estimator $\widehat{\xi}$ for $\xi$ is endowed with a 
mean-squared error $\langle (\widehat{\xi} -\xi)^2 \rangle$ which is lower bounded by the 
Cram\'er-Rao bound involving the reciprocal of the classical Fisher information $F_c(\xi)$. The 
larger the Fisher information $F_c(\xi)$, the more efficient the estimation can be. The maximum 
likelihood estimator \cite{Kay93} is known to achieve the best efficiency dictated by the 
Cram\'er-Rao bound and Fisher information $F_c(\xi)$, at least in the asymptotic regime of a 
large number of independent data points. The classical Fisher information $F_c(\xi)$ stands in 
this respect as a fundamental metric quantifying the best achievable efficiency in estimation. 
When the data are distributed according to the $\xi$-dependent probability distribution
$P_m(\xi)$, the classical Fisher information is defined as
\begin{equation}
F_c(\xi)= \sum_m \dfrac{[\partial_\xi P_m(\xi)]^2}{P_m(\xi)} \;.
\label{Fc_def}
\end{equation}

For estimation from a $\xi$-dependent qubit state $\rho_\xi$ of Bloch vector $\vec{r}_\xi$, a 
useful approach is to perform a spin measurement, which amounts to measuring the observable 
$\vec{\omega}\cdot \vec{\sigma}$ characterized by the unit vector 
$\vec{\omega} \in \mathbbm{R}^3$. Two measurement outcomes follow with the $\xi$-dependent 
probabilities
\begin{equation}
P_\pm (\xi)=\dfrac{1}{2} \bigl(1 \pm \vec{\omega}\, \vec{r}_\xi \bigr)\;,
\label{Pp+__}
\end{equation}
controlled by the scalar inner product $\vec{\omega}\, \vec{r}_\xi $ in $\mathbbm{R}^3$.
The classical Fisher information of Eq.~(\ref{Fc_def}) then follows as
\begin{equation}
F_c(\xi)= \dfrac{(\vec{\omega}\,\partial_\xi \vec{r}_\xi)^2}{1-(\vec{\omega}\, \vec{r}_\xi)^2} \;.
\label{Fc1}
\end{equation}

\bigbreak
It is also possible to obtain further assessment of the performance in quantum estimation, without 
referring to an explicit measurement protocol or measurement vector $\vec{\omega}$. This can be 
accomplished with the quantum Fisher information \cite{Barndorff00,Paris09}. The quantum Fisher 
information is universally used for performance assessment in many areas of quantum metrology, 
with finite-dimensional, or infinite-dimensional, or continuous quantum states 
\cite{Fujiwara95,Gibilisco09,Alipour15,Chapeau17,Birchall20}.
For a $\xi$-dependent qubit state $\rho_\xi$ of Bloch vector $\vec{r}_\xi$, the quantum Fisher 
information relative to the parameter $\xi$ can be expressed \cite{Chapeau16} as
\begin{equation}
F_q(\xi) = \dfrac{\bigl(\vec{r}_\xi\, \partial_\xi\vec{r}_\xi\bigr)^2}{1-\vec{r}_\xi^{\; 2}} +
\bigl( \partial_\xi\vec{r}_\xi \bigr)^2 \;,
\label{Fq_noisyqb}  
\end{equation}
for the general case of a mixed state $\rho_\xi$, while it reduces to
$F_q(\xi) = \bigl( \partial_\xi\vec{r}_\xi \bigr)^2$ for the special case of a pure state 
$\rho_\xi$. The quantum Fisher information $F_q(\xi)$ is intrinsic to the relation of the quantum 
state $\rho_\xi$ to the parameter $\xi$, and does not refer to any measurement performed on 
$\rho_\xi$, but depends only on the functional dependence of $\rho_\xi$ on $\xi$, as visible from 
Eq.~(\ref{Fq_noisyqb}). By contrast, the classical Fisher information $F_c(\xi)$ is determined by 
the probability distribution of the measurement outcomes, as visible in Eq.~(\ref{Fc_def}), and is 
therefore tied to a specific quantum measurement. The usefulness of $F_q(\xi)$ is that it 
constitutes an upper bound to $F_c(\xi)$, imposing $F_c(\xi) \le F_q(\xi)$. There might not 
always exist a fixed $\xi$-independent measurement protocol to achieve $F_c(\xi) = F_q(\xi)$, 
however iterative strategies implementing adaptive measurements
\cite{Barndorff00,Armen02,Fujiwara06,Brivio10,Tesio11,Okamoto12} are accessible to achieve 
$F_c(\xi) = F_q(\xi)$. The quantum Fisher information $F_q(\xi)$ is therefore a meaningful 
metric to characterize the overall best performance for estimation.

\subsection{For the control qubit of the switched channel}

For estimating the phase $\xi$ through the measurement of the control qubit displaying the two 
outcomes characterized by the probabilities $P^{\rm con}_\pm$ of Eq.~(\ref{Pc+a}), the classical 
Fisher information of Eq.~(\ref{Fc_def}) is
\begin{equation}
F_c^{\rm con}(\xi)= \dfrac{(\partial_\xi P^{\rm con}_+)^2}{(1-P^{\rm con}_+)P^{\rm con}_+} \;.
\label{Fc2}
\end{equation}
This form related to Eq.~(\ref{Pc+a}) shows that the most favorable condition to maximize 
$F_c^{\rm con}(\xi)$ of Eq.~(\ref{Fc2}) is to choose $p_c=1/2$, which amounts to preparing the 
control qubit in the state $\ket{\psi_c}=\ket{+}$, and places the switched channel in a maximally 
indefinite causal order; we shall stick to this favorable condition $p_c=1/2$ in the sequel. From
Eqs.~(\ref{Pc+a}) and (\ref{Qa1}), the Fisher information of Eq.~(\ref{Fc2}) then follows as
\begin{eqnarray}
\label{Fc3_b}
F_c^{\rm con}(\xi) &=& \dfrac{\bigl[\partial_\xi Q_\xi(\alpha) \bigr]^2}{1-Q^2_\xi (\alpha)} \\
&=& \dfrac{\bigl[ (1-\alpha)\alpha \sin(\xi) \bigr]^2}
{1-\Bigl[ (1-\alpha)\alpha\cos(\xi)+\dfrac{1}{4}(1+\alpha)^2 \Bigr]^2} \;.
\label{Fc3}
\end{eqnarray}

As already anticipated from Eq.~(\ref{Pc+a}), the performance $F_c^{\rm con}(\xi)$ in 
Eq.~(\ref{Fc3}) upon measuring the control qubit, is independent of the rotation axis $\vec{n}$ 
and of the situation of the input probe $\vec{r}$ specially in relation to $\vec{n}$\,; it is 
obtained uniformly for any probe $\vec{r}$ and axis $\vec{n}$. This would not be the case upon 
measuring a probe qubit in a conventional approach, as we are going to see in the next section.

\bigbreak
Further assessment of the estimation performance is provided by the 
quantum Fisher information of Eq.~(\ref{Fq_noisyqb}). When the control qubit is measured for 
estimating the phase $\xi$ while the probe qubit is left untouched or unobserved, it is possible to 
assign a $\xi$-dependent state $\rho^{\rm con}_\xi$ to the control qubit by tracing over the probe 
qubit in the joint probe-control state $\mathcal{S}(\rho \otimes \rho_c)$ of Eq.~(\ref{Sgenqb}), 
yielding
\begin{eqnarray}
\nonumber
\rho^{\rm con}_\xi=\tr_{\rm probe}\bigl[ \mathcal{S}(\rho \otimes \rho_c) \bigr]
&=& \tr[ \mathcal{S}_{00}(\rho)] \,\bigl[p_c \ket{0}\bra{0}_c + (1-p_c) \ket{1}\bra{1}_c \bigr]\\
\label{Sgenqb_tp1}
&+& \tr[\mathcal{S}_{01}(\rho)] \, \sqrt{(1-p_c)p_c} \bigl( \ket{0}\bra{1}_c + 
\ket{1}\bra{0}_c \bigr) \;.
\end{eqnarray}
From Eq.~(\ref{NUNU_ro}) one has $\tr[ \mathcal{S}_{00}(\rho)] =1$.
From Eq.~(\ref{S01_sr2}) one has $\tr[ \mathcal{S}_{01}(\vec{r}\cdot \vec{\sigma}) ] =0$,
so that $\tr[\mathcal{S}_{01}(\rho)]=\tr[\mathcal{S}_{01}(\mathrm{I}_2)/2]=Q_\xi(\alpha)$
by virtue of Eqs.~(\ref{S01_I2}) and (\ref{Pc+})--(\ref{Qa1}).

The state of the control qubit follows as
\begin{equation}
\rho^{\rm con}_\xi=p_c \ket{0}\bra{0}_c + (1-p_c) \ket{1}\bra{1}_c  + 
Q_\xi(\alpha) \sqrt{(1-p_c)p_c} \bigl( \ket{0}\bra{1}_c + \ket{1}\bra{0}_c \bigr) \;,
\label{Sgenqb_tp2}
\end{equation}
which represents the qubit state characterized by the Bloch vector
$\vec{r}_\xi^{\rm \,con} =\bigl[2\sqrt{(1-p_c)p_c} Q_\xi(\alpha), 0, 2p_c-1 \bigr]^\top$. 
The measurement in the Fourier basis $\bigl\{\ket{+}, \ket{-}\bigr\}$ of the control qubit is 
equivalent to a spin measurement with vector $\vec{\omega}_c =\vec{e}_x=[1, 0, 0]^\top$ acting on 
$\vec{r}_\xi^{\rm \,con}$ via Eq.~(\ref{Pp+__}) to deliver the probabilities $P^{\rm con}_\pm$ of
Eq.~(\ref{Pc+a}). In addition, with the derivative $\partial_\xi \vec{r}_\xi^{\rm \,con} 
=\bigl[2\sqrt{(1-p_c)p_c}\partial_\xi Q_\xi(\alpha), 0, 0 \bigr]^\top$,
Eq.~(\ref{Fq_noisyqb}) readily provides the quantum Fisher information $F_q^{\rm con}(\xi)$
associated with the control qubit. It can then be verified that this $F_q^{\rm con}(\xi)$ is 
maximized at $p_c=1/2$, which provides an additional motivation to this favorable configuration 
for preparing the control qubit. Then at $p_c=1/2$, one has 
$\vec{r}_\xi^{\rm \,con} =[Q_\xi(\alpha), 0, 0]^\top$ and
$\partial_\xi \vec{r}_\xi^{\rm \,con} =[\partial_\xi Q_\xi(\alpha), 0, 0]^\top$, and
Eq.~(\ref{Fq_noisyqb}) gives for the control qubit the quantum Fisher information
\begin{eqnarray}
\label{Fq_tp1}
F_q^{\rm con}(\xi) 
&=& \dfrac{\bigl[Q_\xi(\alpha)  \partial_\xi Q_\xi(\alpha) \bigr]^2}{1-Q^2_\xi (\alpha)} 
+ \bigl[\partial_\xi Q_\xi(\alpha) \bigr]^2 \\
&=& \dfrac{\bigl[\partial_\xi Q_\xi(\alpha) \bigr]^2}{1-Q^2_\xi (\alpha)}  \;,
\label{Fq_tp2}
\end{eqnarray}
which coincides with the classical Fisher information $F_c^{\rm con}(\xi)$ of
Eqs.~(\ref{Fc3_b})--(\ref{Fc3}). This indicates that the measurement protocol chosen for the 
control qubit, which achieves $F_c^{\rm con}(\xi) =F_q^{\rm con}(\xi)$, represents the most 
efficiency strategy for estimating the phase $\xi$ from the control qubit.

\bigbreak
A typical evolution of the Fisher information $F_c^{\rm con}(\xi) =F_q^{\rm con}(\xi)$ from 
Eq.~(\ref{Fc3}) is presented in Fig.~\ref{figFc1}, especially as a function of the level of the 
depolarizing noise quantified by the compression factor $\alpha$. 
At maximum compression at $\alpha =0$, the noise $\mathcal{N}(\cdot)$ in Eq.~(\ref{depol2}) 
completely depolarizes the probe qubit, the measurement probabilities $P^{\rm con}_\pm$ in 
Eq.~(\ref{Pc+a}) for the control qubit become independent of the phase $\xi$, and the Fisher 
information $F_c^{\rm con}(\xi) =F_q^{\rm con}(\xi)$ in Eq.~(\ref{Fc3}) vanishes, indicating that 
at maximum noise the control qubit can no longer serve to estimate $\xi$. 
But also, when there is no noise, at $\alpha =1$, Eq.~(\ref{Pc+a}) shows that the 
measurement probabilities $P^{\rm con}_\pm$ no longer depend on the phase $\xi$, and this entails 
a vanishing Fisher information $F_c^{\rm con}(\xi) =F_q^{\rm con}(\xi)$ in Eq.~(\ref{Fc3}). This 
relates to the observation made at the end of Section~\ref{qbswitch_sec}, that with no noise the 
control qubit does not get coupled to the probe and remains independent of the phase $\xi$, and 
cannot serve to its estimation. 
In between, for intermediate levels of noise with $\alpha \in (0, 1)$, phase estimation from the 
control qubit is possible, as indicated by a non-vanishing Fisher information 
$F_c^{\rm con}(\xi) =F_q^{\rm con}(\xi)$ in Eq.~(\ref{Fc3}). In addition, as illustrated in 
Fig.~\ref{figFc1}, there exists an optimal value of the noise compression factor $\alpha$, 
strictly between $0$ and $1$ (around $\alpha \approx 0.6$ in Fig.~\ref{figFc1}), that maximizes 
the Fisher information $F_c^{\rm con}(\xi) =F_q^{\rm con}(\xi)$ of Eq.~(\ref{Fc3}). This indicates 
that there is in general an optimal nonzero amount of noise to maximize the efficiency of 
estimating the phase $\xi$ from the control qubit of the switched channel. This is reminiscent of 
the phenomenon of stochastic resonance, which characterizes situations where maximum efficiency for 
information processing is obtained at a nonzero level of noise, and which in the quantum context 
assigns a beneficial role to decoherence 
\cite{Gammaitoni98,Chapeau99b,Ting99,Bowen06,Chapeau15c,Gillard17,Gillard19}, 
and relates at a broader level to nontrivial interactions among information, fluctuations and 
noise \cite{Goold16,Zanin19}.

\subsection{Comparison with a standard probe qubit} \label{standqb_sec}

A useful reference is the classical Fisher information for estimating the phase $\xi$ from the 
measurement of a probe qubit that would interact with the noisy unitary channel in a conventional 
one-stage cascade as in Fig.~\ref{figUxiN}, with no quantum switch of the channel. For the probe 
qubit prepared in the state $\rho$ of Eq.~(\ref{roBloch}), one pass through this channel of 
Fig.~\ref{figUxiN} is described by the quantum operation of Eq.~(\ref{NUxi_ro}), and it leaves the 
qubit in a state characterized by the Bloch vector
\begin{equation}
\vec{r}_1(\xi)= \alpha U_\xi \vec{r} \;.
\label{r1}
\end{equation}
Moreover, for a qubit experiencing the quantum process of Eqs.~(\ref{NUxi_ro}) and (\ref{r1}), 
Ref.~\cite{Chapeau16} shows that the derivative 
$\partial_\xi\vec{r}_1 = \vec{n} \times \vec{r}_1$. 
Therefore, with $\vec{r}_1(\xi)\equiv \vec{r}_\xi$ placed in Eq.~(\ref{Fc1}) one obtains the 
Fisher information
\begin{equation}
F_c(\xi)=\dfrac{[\vec{\omega}( \vec{n} \times \vec{r}_1 )]^2}
{1-(\vec{\omega} \vec{r}_1)^2}  =  \dfrac{\alpha^2[\vec{\omega}( \vec{n} \times U_\xi \vec{r}\,)
]^2} {1-\alpha^2(\vec{\omega} U_\xi \vec{r}\,)^2} \;,
\label{Fc1_p}
\end{equation}
upon measuring the qubit spin observable $\vec{\omega}\cdot \vec{\sigma}$. 

Equation~(\ref{Fc1_p}) shows that the Fisher information $F_c(\xi)$, and therefore the maximum 
performance in estimating $\xi$, is strongly dependent on the situation in $\mathbbm{R}^3$ of the
input probe $\vec{r}$ and of the measurement vector $\vec{\omega}$ in relation to the rotation 
axis $\vec{n}$. As analyzed for instance in Ref.~\cite{Chapeau16}, maximizing the Fisher 
information $F_c(\xi)$ of Eq.~(\ref{Fc1_p}) requires a pure input probe with $\vec{r}$ orthogonal 
to the rotation axis $\vec{n}$\,; in addition it requires a measurement vector $\vec{\omega}$ 
orthogonal to both the axis $\vec{n}$ and the rotated Bloch vector $\vec{r}_1(\xi)$. When these 
conditions are satisfied, Eq.~(\ref{Fc1_p}) reaches the overall maximum $F_c^{\rm max}(\xi)=\alpha^2$.
This maximum can hardly be generally reached in practice since in particular satisfying
$\vec{\omega} \perp \vec{r}_1(\xi)$ would require to know the rotation angle $\xi$ under estimation.
By contrast, the control qubit of the switched channel uniformly reaches the performance 
$F_c^{\rm con}(\xi)$ of Eq.~(\ref{Fc3}), for any input probe $\vec{r}$ (pure or mixed) and
with a fixed measurement vector $\vec{\omega}_c =\vec{e}_x$.

\begin{figure}[htb]
{\includegraphics[width=84mm]{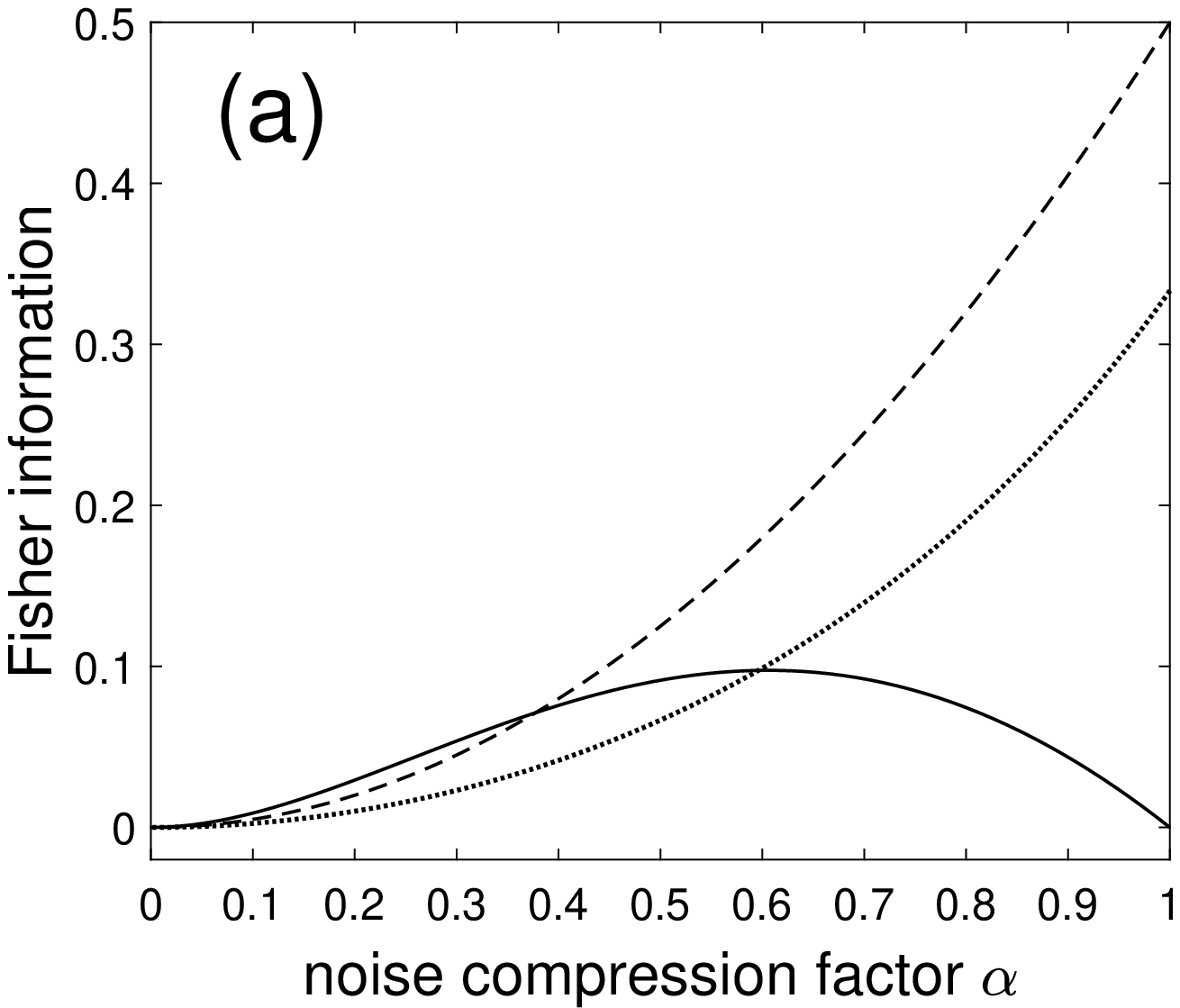}}
\hfill
{\includegraphics[width=84mm]{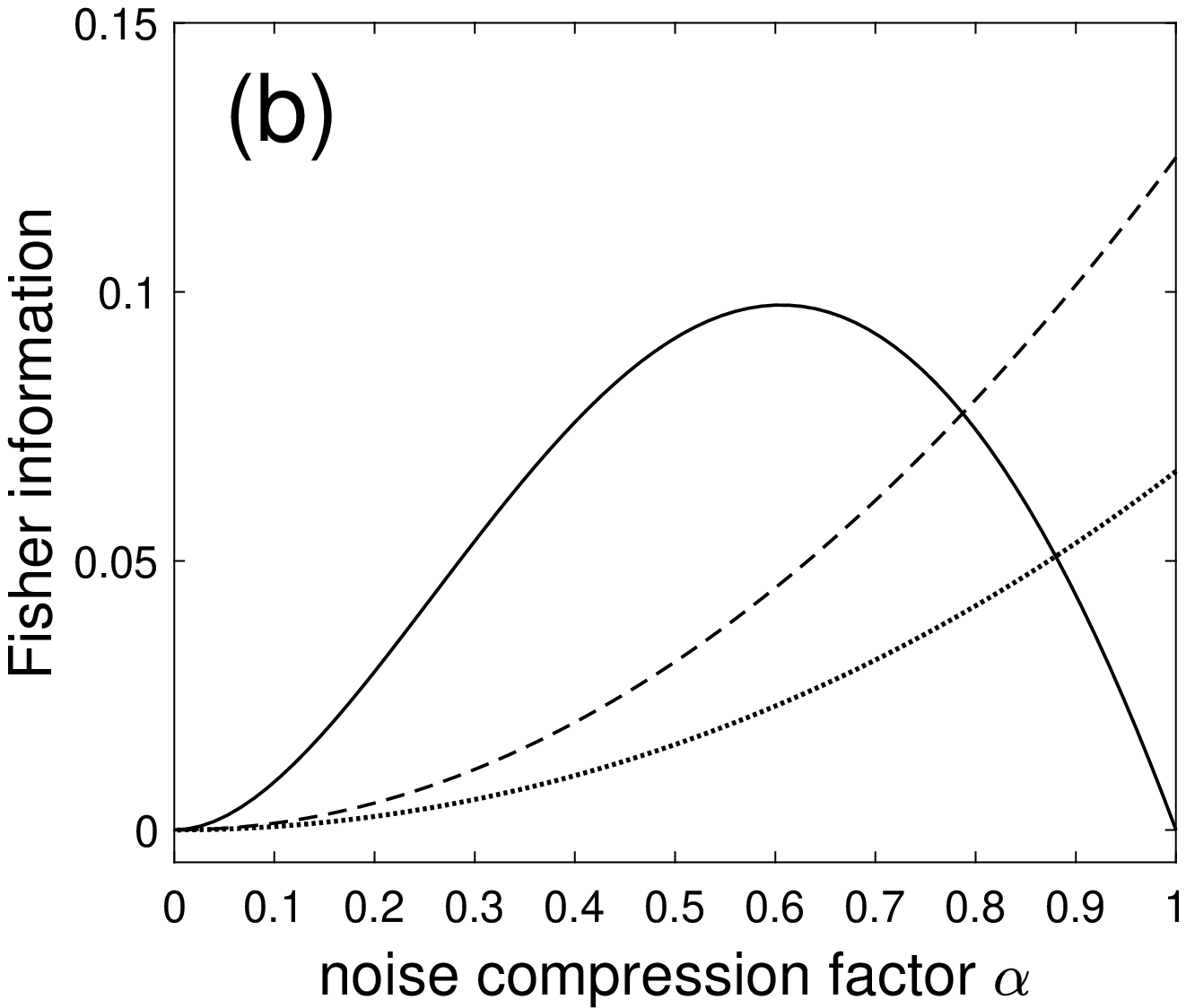}}
\caption[what appears in lof LL p177]
{The unitary transformation $\mathsf{U}_\xi$ of Eq.~(\ref{Uxi}) is with axis 
$\vec{n}=[1, 0, 1]^\top / \sqrt{2}$ and phase angle $\xi =\pi /2$. 
As a function of the compression factor $\alpha$ of the depolarizing noise of Eq.~(\ref{depol2}),
the solid line is the Fisher information $F_c^{\rm con}(\xi) =F_q^{\rm con}(\xi)$ from 
Eq.~(\ref{Fc3}) upon measuring the control qubit of the switched quantum channel. The dotted line 
is the classical Fisher information $F_c(\xi)$ of Eq.~(\ref{Fc1_p}) upon measuring a probe qubit 
after a one-stage standard cascade as in Fig.~\ref{figUxiN} with measurement Bloch vector 
$\vec{\omega}=\vec{e}_x$, which is upper-bounded by the quantum Fisher information $F_q(\xi)$ of 
Eq.~(\ref{Fpur_qb}) represented by the dashed line.
In (a) the input probe of Eq.~(\ref{roBloch}) is in the pure state $\rho$ with unit Bloch vector 
$\vec{r}=[1, 0, 0]^\top =\vec{e}_x$; in (b) the input probe is in the mixed state $\rho$ with 
$\vec{r}=[0.5, 0, 0]^\top =\vec{e}_x/2$.
}
\label{figFc1}
\end{figure}

A typical evolution of the classical Fisher information $F_c(\xi)$ of Eq.~(\ref{Fc1_p}) is shown 
in Fig.~\ref{figFc1}, in a configuration where $\vec{r}$ and $\vec{\omega}$ are not optimized as 
orthogonal to $\vec{n}$ (supposedly because $\vec{n}$ is not precisely known), and for comparison 
with the situation of the control qubit of Eq.~(\ref{Fc3}) which is insensitive to $\vec{n}$.

\bigbreak 

For a qubit experiencing the quantum process of Eq.~(\ref{NUxi_ro}), as indicated above, one has 
$\partial_\xi\vec{r}_\xi = \vec{n} \times \vec{r}_\xi$ so that 
$\vec{r}_\xi\, \partial_\xi\vec{r}_\xi=0$; with $\vec{r}_\xi \equiv \vec{r}_1(\xi)$ from 
Eq.~(\ref{r1}), the quantum Fisher information of Eq.~(\ref{Fq_noisyqb}) then reduces to
\begin{equation}
F_q(\xi) = \bigl( \partial_\xi\vec{r}_\xi \bigr)^2 =\alpha^2 (\vec{n}\times\vec{r} \,)^2\;.
\label{Fpur_qb}
\end{equation}  
The overall maximum $F_q^{\rm max}(\xi)=\alpha^2$ is achieved in Eq.~(\ref{Fpur_qb}) with a 
unit-norm input Bloch vector $\vec{r}$ orthogonal to the rotation axis $\vec{n}$.
A typical evolution of $F_q(\xi)$ of Eq.~(\ref{Fpur_qb}) is also presented in Fig.~\ref{figFc1}.

\smallbreak

\bigbreak 

When the Bloch vector $\vec{r}$ of the input probe tends to align with the rotation axis $\vec{n}$, 
Eq.~(\ref{Fpur_qb}) shows that the quantum Fisher information $F_q(\xi)$ tends to vanish, and so 
does any classical Fisher information $F_c(\xi)$ attached to any measurement protocol of the probe 
qubit (even generalized measurements). In this circumstance, measurement of the probe qubit involved 
in the conventional approach of Eqs.~(\ref{NUxi_ro}) and (\ref{r1}) becomes inefficient to estimate 
the rotation angle $\xi$. By contrast, as with the measurement probabilities $P^{\rm con}_\pm$ of 
Eq.~(\ref{Pc+a}), the Fisher information $F_c^{\rm con}(\xi)=F_q^{\rm con}(\xi)$ of Eq.~(\ref{Fc3}) 
for the control 
qubit of the switched channel, is unaffected by the Bloch vector $\vec{r}$ of the input probe and 
the axis $\vec{n}$ of the unitary $\mathsf{U}_\xi$. As a result, measurement of the control qubit 
of the switched channel keeps the same estimation efficiency of Eq.~(\ref{Fc3}) irrespective of 
the situation of the input Bloch vector $\vec{r}$ in relation to the rotation axis $\vec{n}$. The 
rotation by $\xi$ can even take place on a probe vector $\vec{r}$ parallel to the rotation axis 
$\vec{n}$, and with $\vec{r}\varparallel \vec{n}$ the control qubit keeps the same estimation 
efficiency of Eq.~(\ref{Fc3}), while the conventional approach of Eqs.~(\ref{NUxi_ro}) and (\ref{r1}) 
becomes inoperative.

In a comparable way, when the input probe depolarizes as $\| \vec{r}\, \| \rightarrow 0$,
the conventional approach of Eqs.~(\ref{NUxi_ro}) and (\ref{r1}) gradually loses its efficiency 
for estimating the phase $\xi$, as marked by $F_q(\xi)$ in Eq.~(\ref{Fpur_qb}) which vanishes as 
$\| \vec{r}\, \| \rightarrow 0$. By contrast, the switched channel via its control qubit
keeps the same estimation efficiency as in Eq.~(\ref{Fc3}), for any $\| \vec{r}\, \|$.
Even at $\| \vec{r}\, \| =0$, when probing with a fully depolarized input probe in the maximally 
mixed state $\rho=\mathrm{I}_2/2$ in Eq.~(\ref{roBloch}), the control qubit of the switched channel 
remains equally efficient for estimation, while the conventional approach of Eqs.~(\ref{NUxi_ro}) 
and (\ref{r1}) becomes inoperative.

Figure~\ref{figFc1} illustrates in particular the impact of an input probe $\vec{r}$ not
orthogonal to the rotation axis $\vec{n}$. 
At low noise, when the compression factor $\alpha$ is close to $1$ in Fig.~\ref{figFc1},
direct estimation from the standard cascade is more efficient. 
Yet, as the level of noise increases when $\alpha$ approaches $0$, the performance of the 
control qubit of the switched channel quantified by $F_c^{\rm con}(\xi)=F_q^{\rm con}(\xi)$ of 
Eq.~(\ref{Fc3}), gradually outperforms both $F_c(\xi)$ and then $F_q(\xi)$ characterizing the 
standard cascade of Fig.~\ref{figUxiN} in conventional estimation.
With decreasing $\| \vec{r}\, \|$, when passing from a pure input probe in Fig.~\ref{figFc1}(a) to 
a mixed input probe in Fig.~\ref{figFc1}(b), the performance of the control qubit is unaffected, 
while the performance of the standard cascade is reduced. This advantage of the control qubit 
would get more pronounced and would occur earlier (for $\alpha$ closer to $1$) as the input probe 
$\vec{r}$ approaches the axis $\vec{n}$ or shrinks as $\| \vec{r}\, \| \rightarrow 0$, as 
explained above. This illustrates the regime of interest for qubit metrology, with an
ill-configured input probe $\vec{r}$ or for blind estimation with an unknown axis $\vec{n}$,
when the control qubit of the switched channel maintains a uniform unaffected efficiency,
while conventional estimation in the standard cascade becomes less efficient.

\subsection{Comparison with a two-stage standard cascade} \label{standqb2_sec}

Although the control qubit never directly interacts with the unitary $\mathsf{U}_\xi$ under
estimation, the switched channel involves two passes of its probe qubit across the unitary 
$\mathsf{U}_\xi$, and it may be compared with a two-stage cascade of a conventional estimation.
Equation~(\ref{Fc1_p}) gives also access to the characterization of a two-stage cascading of the 
noisy unitary channel involved in Eq.~(\ref{r1}), in a standard way with definite causal order.
Instead of the one-stage cascading acting as 
$\vec{r}\mapsto \alpha U_\xi \vec{r}$ via Eq.~(\ref{r1}), the two-stage cascading acts as 
$\vec{r}\mapsto \alpha^2 U_\xi^2 \vec{r} =\alpha^2 U_{2\xi}\vec{r}$ and is therefore 
equivalent to a one-stage cascading with the rotation angle $2\xi$ instead of $\xi$ at a noise 
compression $\alpha^2$ instead of $\alpha$. Through these two changes, if the one-stage classical 
Fisher information of Eq.~(\ref{Fc1_p}) is denoted $F_c(\xi, \alpha)$ then the two-stage cascading 
is characterized by the Fisher information $F_c(2\xi, \alpha^2)$. The one-stage cascading can 
provide an estimation of the angle $\xi$ with a minimal root-mean squared (rms) error evolving as 
$\sim 1/\sqrt{F_c(\xi, \alpha)}$; meanwhile the two-stage cascading can provide an estimation of 
the angle $2\xi$ with a minimal rms error evolving as $\sim 1/\sqrt{F_c(2\xi, \alpha^2)}$, which
provides an estimation for $\xi$ with the halved rms error $\sim 1/[2\sqrt{F_c(2\xi, \alpha^2)}]$. 
For an assessment of a conventional estimation of $\xi$, it is therefore meaningful to confront
$F_c(\xi, \alpha)$ and $4F_c(2\xi, \alpha^2)$: for a given $\xi$ and a given noise compression 
$\alpha$, one measurement of the probe qubit after the one-stage cascade delivers about $\xi$ a 
Fisher information $F_c(\xi, \alpha)$, while one measurement of the probe qubit after the 
two-stage cascade delivers about $\xi$ a Fisher information $4F_c(2\xi, \alpha^2)$. 
The two-stage cascade amplifies by $2$ the parameter $\xi$ to be estimated, entailing a reduced 
error, but is also more exposed to the noise, compared with the one-stage cascade. As a result, 
typically, it can 
be observed that $4F_c(2\xi, \alpha^2)$ is superior to $F_c(\xi, \alpha)$ at small compression 
with $\alpha$ close to $1$, indicating that the two-stage cascade is more efficient for estimating 
$\xi$ at low noise level; meanwhile, $F_c(\xi, \alpha)$ is superior to $4F_c(2\xi, \alpha^2)$ at 
large compression with $\alpha$ close to $0$, indicating that the one-stage cascade is more 
efficient for estimating $\xi$ at high noise level.
A similar picture is conveyed by the quantum Fisher information $F_q(\xi)$ of Eq.~(\ref{Fpur_qb}), 
with $F_q(\xi, \alpha) =\alpha^2 (\vec{n}\times\vec{r} \,)^2$ for the one-stage cascade, to be 
confronted with $4F_q(2\xi, \alpha^2) =4\alpha^4 (\vec{n}\times\vec{r} \,)^2$ for the two-stage 
cascade, with the first which is superior at high noise level (small $\alpha$), and which is used 
for comparison with the switched channel in Figs.~\ref{figFq1}--\ref{figFq3}.

\subsection{Phase-averaged performance}

\begin{figure}[htb]
\centerline{\includegraphics[width=84mm]{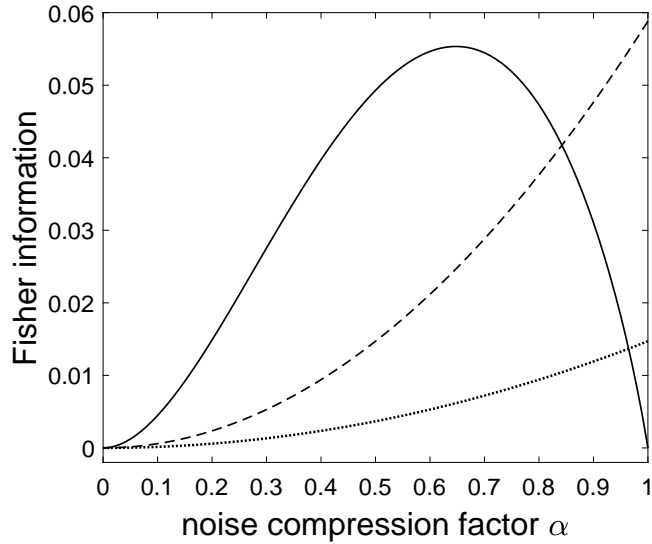}}
\caption[what appears in lof LL p177]
{The unitary transformation $\mathsf{U}_\xi$ of Eq.~(\ref{Uxi}) is with axis 
$\vec{n}=[0.8, 0, 0.2]^\top / \sqrt{0.68}$.
As a function of the compression factor $\alpha$ of the depolarizing noise of Eq.~(\ref{depol2}),
the solid line is the Fisher information $\overline{F}_c^{\rm con}$ of Eq.~(\ref{Fc_av}) after 
averaging over $\xi$ of $F_c^{\rm con}(\xi)=F_q^{\rm con}(\xi)$ of Eq.~(\ref{Fc3}) for the control 
qubit of the switched channel. The dashed line is the quantum Fisher information $F_q(\xi)$ of 
Eq.~(\ref{Fpur_qb}) for a one-stage standard cascade as in Fig.~\ref{figUxiN} and when the input 
probe in Eq.~(\ref{roBloch}) is in the pure state with unit Bloch vector 
$\vec{r}=[1, 0, 0]^\top =\vec{e}_x$, the dotted line is $F_q(\xi)$ of Eq.~(\ref{Fpur_qb}) for the 
mixed input probe with $\vec{r}=[0.5, 0, 0]^\top =\vec{e}_x/2$.
}
\label{figFc2}
\end{figure}

The classical Fisher information, $F_c(\xi)$ for a standard probe qubit in Eq.~(\ref{Fc1_p}), 
or $F_c^{\rm con}(\xi)$ 
for the control qubit of the switched quantum channel in Eq.~(\ref{Fc3}), is dependent on the 
phase angle $\xi$. This is a common property, often observed for quantum phase estimation in the 
presence of noise, and implying a performance varying according to the range of the phase $\xi$ to 
be estimated. A measurement result depending on $\xi$ is necessary to enable estimation of $\xi$ 
by such measurement. Commonly this entails also a measurement performance depending on $\xi$, 
and related here to the geometric configuration of the rotated Bloch vector $U_\xi \vec{r}$ in 
$\mathbbm{R}^3$. For assessing the performance, it can be meaningful to consider the averaged 
Fisher information $\overline{F}_c = \int_0^{2\pi}F_c(\xi) d\xi/(2\pi)$ reflecting the average 
performance for values of $\xi$ uniformly covering the interval $[0, 2\pi )$.
Especially, for the Fisher information $F_c^{\rm con}(\xi)$ in Eq.~(\ref{Fc3}) of the control 
qubit of the switched channel, the integral over $\xi$ can be worked out explicitly to give
\begin{equation}
\overline{F}_c^{\rm con} = 1-\dfrac{\sqrt{3}}{8}(1-\alpha) \sqrt{(1-\alpha)(3+5\alpha)} 
 - \dfrac{1}{8} \sqrt{(5+6\alpha-3\alpha^2)(5-2\alpha+5\alpha^2)} \;.
\label{Fc_av}
\end{equation}  
The average Fisher information in Eq.~(\ref{Fc_av}) especially satisfies
$\overline{F}_c^{\rm con}(\alpha =0)=\overline{F}_c^{\rm con}(\alpha =1)=0$ as expected. It is 
represented and compared in Fig.~\ref{figFc2} in conditions where the control qubit of the
switched channel offers useful specific capabilities for estimation, with an input probe 
$\vec{r}$ tending to align with the axis $\vec{n}$ or with a mixed input probe of
$\| \vec{r}\, \| <1$.

\smallbreak

The average Fisher information $\overline{F}_c^{\rm con}$, in the conditions of Fig.~\ref{figFc2}, 
illustrates that on average over the whole range $[0, 2\pi )$ of the phase $\xi$, the control qubit 
of the switched channel can offer, as the level of noise increases, higher efficiency for estimation 
compared with the standard cascade of Fig.~\ref{figUxiN} having to operate with a non-optimized
input probe. And this advantage would get more pronounced as the input probe $\vec{r}$ continues to 
approach the axis $\vec{n}$ or further depolarizes as $\| \vec{r}\, \| \rightarrow 0$.

\subsection{Comparison with two qubits in conventional estimation} \label{2convqb_sec}

So far, in the switched channel, measurement of the control qubit, which never directly interacts 
with the unitary $\mathsf{U}_\xi$ under estimation, has been compared with measurement of a qubit 
involved in a conventional estimation process, in a one-stage or two-stage interaction with the 
unitary $\mathsf{U}_\xi$. This amounts to comparing, for estimation, protocols performing a single 
measurement on a single qubit, in the switched channel or in a conventional setting. It can also 
be considered that the switched channel is a two-qubit process, with a control qubit and a probe 
qubit, so that a comparison with two qubits involved in conventional estimation could offer another
meaningful reference. We essentially show below that when estimation has to cope with ill-configured 
input probes, measurement of the control qubit of the switched channel still displays its 
useful specificities compared with conventional techniques involving two or more qubits.

If two independent qubits are measured when repeating the conventional estimation of 
Section~\ref{standqb_sec}, then the Fisher informations $F_c(\xi)$ and $F_q(\xi)$ are additive. 
Especially, from Eq.~(\ref{Fpur_qb}), the quantum Fisher information while measuring two such 
independent probe qubits is $F_q^{(2)}(\xi) = 2\alpha^2 (\vec{n}\times\vec{r} \,)^2$. As 
previously, $F_q^{(2)}(\xi) = 2\alpha^2 (\vec{n}\times\vec{r} \,)^2$ is maximized when the input 
probe qubits are prepared in a pure state with $\| \vec{r}\, \|=1$. 
For an input probe with $\vec{r}$ tending to align with the axis $\vec{n}$ of the unitary 
$\mathsf{U}_\xi$, the conventional estimation measuring two qubits becomes inoperative with 
$F_q^{(2)}(\xi) \rightarrow 0$, while a single measurement of the control qubit of the switched 
channel keeps the same efficiency as in Eq.~(\ref{Fc3}) or (\ref{Fq_tp2}), irrespective of the 
orientation of the input probe. In a similar way, for a depolarized input probe with
$\| \vec{r}\, \| \rightarrow 0$, the conventional estimation measuring two qubits becomes 
inoperative with $F_q^{(2)}(\xi) \rightarrow 0$, while a single measurement of the control qubit
keeps the same efficiency as in Eq.~(\ref{Fc3}) or (\ref{Fq_tp2}), unaffected by the depolarization 
of the probe. 

More generally, a two-qubit conventional estimation can use a two-qubit probe prepared in an 
entangled state $\rho$, and operate according to the various schemes inventoried for instance in 
\cite{Demkowicz14}. For such two-qubit schemes, the quantum Fisher information $F_q^{(2)}(\xi)$, 
due to its general convexity property \cite{Alipour15,Chapeau17}, is again maximized by a pure input 
probe state $\rho=\ket{\psi}\bra{\psi}$. A direct interaction of two entangled probe qubits in 
state $\rho$ with the unitary $\mathsf{U}_\xi$ under estimation would involve the transformation 
$\rho \mapsto 
\mathsf{U}_\xi^{(2)} \rho \mathsf{U}_\xi^{(2)}\mbox{}^\dagger$, with the two-qubit unitary
$\mathsf{U}_\xi^{(2)}=\mathsf{U}_\xi\otimes\mathsf{U}_\xi$ for two active qubits in the probe, or
$\mathsf{U}_\xi^{(2)}=\mathsf{U}_\xi\otimes\mathrm{I}_2$ for one inactive ancilla qubit in the 
probe.  If the qubit unitary under estimation in Eq.~(\ref{Uxi}) is written
$\mathsf{U}_\xi=\exp\bigl(-i\xi \mathsf{G}\bigr)$, with the Hermitian generator
$\mathsf{G} =\vec{n} \cdot \vec{\sigma}/2$, then the resulting two-qubit unitary can be put under 
the form $\mathsf{U}_\xi^{(2)}=\exp\bigl(-i\xi \mathsf{G}_2\bigr)$, with the Hermitian generator 
$\mathsf{G}_2=\mathsf{G}\otimes\mathrm{I}_2 + \mathrm{I}_2\otimes \mathsf{G}$ for two active probe 
qubits, or $\mathsf{G}_2=\mathsf{G} \otimes \mathrm{I}_2$ for one inactive ancilla qubit in the 
probe. In the noise-free case, with a pure two-qubit input probe $\rho=\ket{\psi}\bra{\psi}$, the
quantum Fisher information can be expressed \cite{Paris09,Toth14} as
\begin{equation}
F_q^{(2)}(\xi) = 4 \Bigl( \braket{\psi | \mathsf{G}_2^2 | \psi}-
\braket{\psi | \mathsf{G}_2 | \psi}^2 \Bigr)
=4 \braket{\psi | \Delta\mathsf{G}_2^2 | \psi} \;,
\label{Fq_Gpur}
\end{equation}
with the Hermitian variation operator 
$\Delta\mathsf{G}_2=\mathsf{G}_2 - \braket{\psi | \mathsf{G}_2 | \psi}
\mathrm{I}_2\otimes\mathrm{I}_2$. In the presence of the noise $\mathcal{N}(\cdot)$ affecting the 
unitary $\mathsf{U}_\xi$ as in Fig.~\ref{figUxiN}, the quantum Fisher information is upper bounded 
\cite{Paris09,Toth14} as $F_q^{(2)}(\xi) \le 4 \braket{\psi | \Delta\mathsf{G}_2^2 | \psi}$.
As a result, when the state $\ket{\psi}\in \mathcal{H}_2\otimes \mathcal{H}_2$ of the two-qubit
input probe approaches one of the eigenstates of the Hermitian generator $\mathsf{G}_2$ (which are 
the same as the eigenstates of the two-qubit unitary $\mathsf{U}_\xi^{(2)}$), the Fisher information 
$F_q^{(2)}(\xi)$ tends to vanish, as its upper limit $4 \braket{\psi | \Delta\mathsf{G}_2^2 | \psi}$ 
does. This condition encompasses entangled states $\ket{\psi}$ of the two-qubit input probe.
The Hermitian generator $\mathsf{G} =\vec{n} \cdot \vec{\sigma}/2$ has the two eigenvalues $\pm 1/2$ 
associated with the two eigenstates $\ket{u_\pm}$ determined by $\vec{n}$. It results that 
$\mathsf{U}_\xi=\exp\bigl(-i\xi \mathsf{G}\bigr)$ has the two eigenvalues $\exp(\pm i \xi /2)$ 
associated with the same two eigenstates $\ket{u_\pm}$.
As a consequence, $\mathsf{U}_\xi^{\otimes 2}$ has
the four eigenstates $\bigl\{ \ket{u_+}\otimes\ket{u_+}, \ket{u_-}\otimes\ket{u_-},
\ket{u_+}\otimes\ket{u_-}, \ket{u_-}\otimes\ket{u_+} \bigr\}$, associated respectively with the
four eigenvalues $\bigl\{ e^{i\xi}, e^{-i\xi}, +1, +1 \bigr\}$.
Because of the degeneracy of the eigenvalue $+1$, the two states
$\ket{w_\pm} = \bigl(\ket{u_+}\otimes\ket{u_-} \pm \ket{u_-}\otimes\ket{u_+} \bigr)/\sqrt{2}$ are
two eigenstates of $\mathsf{U}_\xi^{\otimes 2}$ with eigenvalue $+1$, and are two
entangled states of the two qubits. Any linear combination of $\ket{w_+}$ and $\ket{w_-}$
will also in general represent an entangled eigenstate of $\mathsf{U}_\xi^{\otimes 2}$ with
the eigenvalue $+1$. This represents a two-dimensional subspace of $\mathcal{H}_2^{\otimes 2}$.
When such two-qubit entangled states are used to probe the unitary $\mathsf{U}_\xi$
(or when the two-qubit state tends to align with such subspace), they also become 
inoperative for conventional estimation.
These configurations stand 
as the two-qubit analogue of the one-qubit probe tending to align with the axis 
$\vec{n}$  of the unitary $\mathsf{U}_\xi$. So with such two-qubit input probes tending to align 
with $\mathsf{U}_\xi$ in this sense, the conventional estimation measuring two qubits becomes 
inoperative with $F_q^{(2)}(\xi) \rightarrow 0$, while again a single measurement of the control 
qubit of the switched channel keeps the same efficiency as in Eq.~(\ref{Fc3}) or (\ref{Fq_tp2}), 
irrespective of the configuration of the input probe. 

In addition, a fully depolarized two-qubit input probe $\rho=\mathrm{I}_2\otimes\mathrm{I}_2/4$ 
gives the interaction 
$\mathsf{U}_\xi^{(2)} \rho \mathsf{U}_\xi^{(2)}\mbox{}^\dagger =\mathrm{I}_2\otimes\mathrm{I}_2/4$
leaving the probe invariant and insensitive to the phase $\xi$. The corresponding Fisher
information is $F_q^{(2)}(\xi) =0$, and the conventional estimation measuring two qubits becomes 
inoperative with such a depolarized input probe, while again a single measurement of the control 
qubit of the switched channel keeps the same efficiency as in Eq.~(\ref{Fc3}) or (\ref{Fq_tp2}), 
unaffected by the depolarization of the probe. 

Two-qubit conventional estimation can also be considered with only one depolarized qubit
in the two-qubit probe. The switched channel uses a coherent control qubit that does not 
directly interact with the unitary $\mathsf{U}_\xi$ but gets entangled to the probe qubit 
interacting with $\mathsf{U}_\xi$. In this respect, we can consider conventional estimation with 
an input qubit pair of an active qubit and an inactive qubit, prepared in an arbitrary entangled 
(possibly optimized in some way) state. Then, before it can interact with the unitary 
$\mathsf{U}_\xi$ to probe it, we consider that the active qubit gets (or tends to be) completely 
depolarized, by some noise affecting the preparation, and so as to place it in the situation of 
the completely depolarized probe qubit of the switched channel; meanwhile the entangled inactive 
qubit remains untouched and unaffected by the depolarization. Then, in this condition also, the 
two-qubit probe becomes completely inoperative for conventional estimation.
This follows directly by the action of $\mathsf{U}_\xi$ on a fully depolarized qubit, leaving 
its state invariant and insensitive to the phase $\xi$; or also by explicit evaluation of the 
quantum Fisher information obtained for instance from \cite{Chapeau17} for this two-qubit 
estimation scheme. Meanwhile, as indicated, a single measurement of the control qubit of the
switched channel remains operative for estimation when probing with a fully depolarized probe 
qubit.

The above features related to conventional estimation carry over with additional $\xi$-independent 
unitaries intervening in the processing of the two qubits, and also with conventional techniques 
employing and measuring more than two qubits, active or inactive, for multiple probing of the 
unitary $\mathsf{U}_\xi$. With a multiple-qubit input probe tending to align with the unitary or 
approaching a fully depolarized preparation, measurement of the multiple qubits becomes 
inoperative in conventional estimation of $\xi$, while a single measurement of the control qubit 
of the switched channel keeps the same estimation efficiency as in Eq.~(\ref{Fc3}) or 
(\ref{Fq_tp2}).

This of course does not alter the fact that conventional estimation, especially with multiple
qubits, remains useful in its own right, especially in controlled conditions where it can be 
optimized. This, in particular, usually requires the knowledge of the axis $\vec{n}$ of the unitary 
$\mathsf{U}_\xi$ under estimation, with well controlled probing inputs, and conditions exist,
especially in the presence of noise, where the optimal configurations are also dependent on the 
unknown phase itself and therefore inaccessible in practice \cite{Barndorff00,Chapeau16}. As a 
complement, in other distinct yet meaningful conditions that we report here, the switched channel 
can bring additional capabilities for estimation, via the novel approach of switched indefinite 
causal order shown here applicable for qubit phase estimation.

\subsection{For the probe qubit of the switched channel} \label{FiProbe_sec} 

In the switched quantum channel, after measurement of the control qubit, the probe qubit gets 
placed in the conditional state $\rho_\pm^{\rm post}$ of Eqs.~(\ref{ro_post})--(\ref{r_post}) 
which in general also depends on the phase $\xi$. Measuring $\rho_\pm^{\rm post}$ can therefore 
provide useful additional information to estimate $\xi$. For an assessment, it is possible to evaluate 
the Fisher information, classical or quantum, in the state $\rho_\pm^{\rm post}$ about $\xi$. For 
a viewpoint not attached to a specific measurement protocol of $\rho_\pm^{\rm post}$, one can 
turn to the quantum Fisher information of Eq.~(\ref{Fq_noisyqb}) applied to the Bloch 
vector $\vec{r}_\pm^{\rm \,post}(\xi) \equiv \vec{r}_\xi$ from Eq.~(\ref{r_post}). For this 
purpose, $\vec{r}_\pm^{\rm \,post}(\xi) $ in Eq.~(\ref{r_post}) especially involves a quadratic 
function of the matrix $U_\xi$, and from Eq.~(\ref{r_post}) it is feasible to analytically compute 
the derivative $\partial_\xi\vec{r}_\pm^{\rm \,post}(\xi)$, noting that for two $\xi$-dependent 
matrices $A_\xi$ and $B_\xi$ one has the derivative 
$\partial_\xi (A_\xi B_\xi) =(\partial_\xi A_\xi) B_\xi + A_\xi (\partial_\xi B_\xi)$. It is in
this way feasible to obtain an analytical expression for the quantum Fisher information 
$F_q^{\rm swi}(\xi)$ for the probe qubit of the switched channel, from Eq.~(\ref{Fq_noisyqb}), when 
$\vec{r}_\xi \equiv \vec{r}_\pm^{\rm \,post}(\xi)$ from
Eq.~(\ref{r_post}); but this expression is rather bulky and we will not write it here.

\begin{figure}[htb]
{\includegraphics[width=83mm]{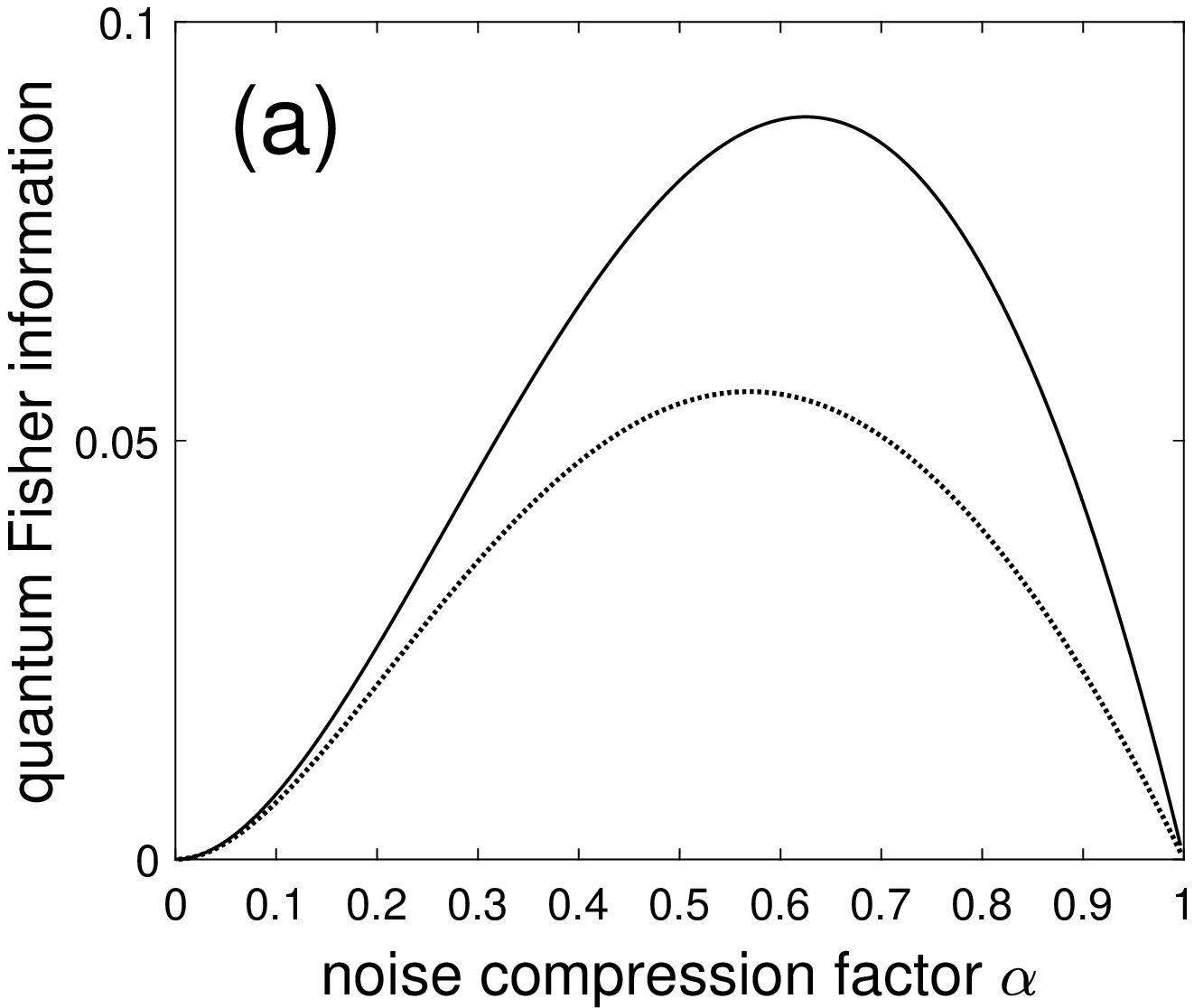}} \hfill
{\includegraphics[width=83mm]{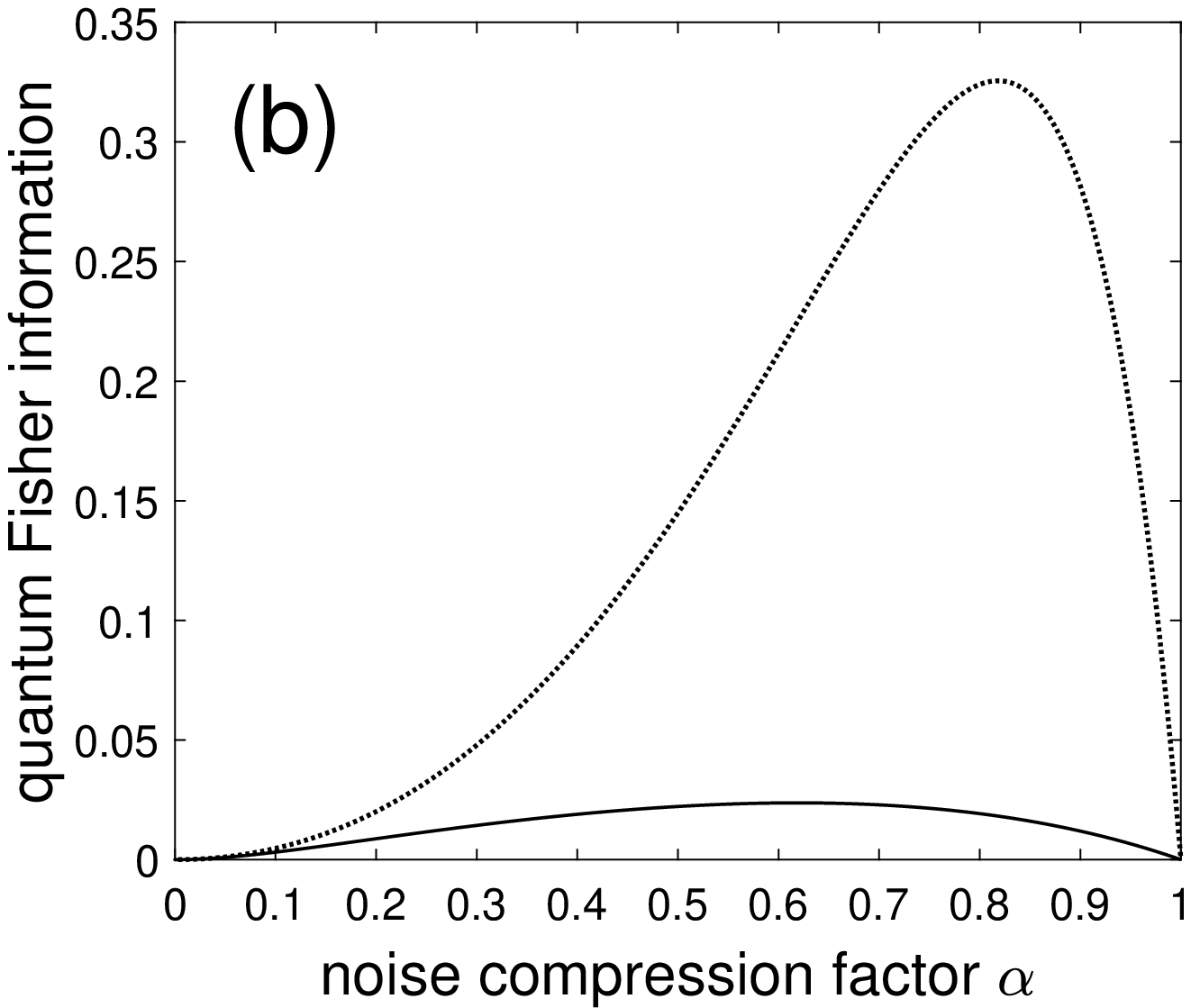}}
\caption[what appears in lof LL p177]
{The unitary transformation $\mathsf{U}_\xi$ of Eq.~(\ref{Uxi}) is with  axis 
$\vec{n}=[1, 0, 0]^\top =\vec{e}_x$, the input probe in Eq.~(\ref{roBloch}) is in the pure state 
with unit Bloch vector $\vec{r}=\vec{n}$, and in abscissa is the compression factor $\alpha$ of the 
depolarizing noise of Eq.~(\ref{depol2}). For the probe qubit of the switched channel, the quantum 
Fisher information $F_q^{\rm swi}(\xi)$ from Eq.~(\ref{Fq_noisyqb}) associated with the state 
$\rho_+^{\rm post}$ (solid line), and $\rho_-^{\rm post}$ (dotted line).
The phase angle is $\xi=\pi /2$ in (a), and $\xi=\pi /4$ in (b).
}
\label{figFq1swi}
\end{figure}

It is observed with the switched channel that in general the Fisher information 
$F_q^{\rm swi}(\xi)$ of the probe states $\rho_+^{\rm post}$ and $\rho_-^{\rm post}$ does depend 
on both $\xi$ and $\vec{n}$, and also on the input probe $\vec{r}$.
As the input probe depolarizes, with $\| \vec{r}\, \| \rightarrow 0$, Eq.~(\ref{r_post}) shows
that $\vec{r}_\pm^{\rm \,post}(\xi)$ both vanish and so does the corresponding Fisher information 
$F_q^{\rm swi}(\xi)$. Therefore, in contrary to the control qubit, the probe qubit of the
switched channel becomes inoperative for estimation with a fully depolarized input probe.
However, in similarity with the control qubit, the estimation efficiency of the probe qubit 
assessed by $F_q^{\rm swi}(\xi)$ does not uniformly vanish for an input probe $\vec{r}$ parallel 
to the rotation axis $\vec{n}$. This is a specific property of the switched channel for estimation, 
accessible both by measuring either the control qubit or the probe qubit, enabling to perform 
estimation of the phase angle $\xi$ even with an input probe $\vec{r}$ parallel to the axis 
$\vec{n}$ of the unitary $\mathsf{U}_\xi$. This property is not present in conventional estimation,
as addressed in Sections \ref{standqb_sec} and \ref{standqb2_sec}, even when repeated on multiple 
qubits as in Section \ref{2convqb_sec}. This specific property of the probe qubit of the switched 
channel is illustrated in Fig.~\ref{figFq1swi}, showing a nonzero Fisher information 
$F_q^{\rm swi}(\xi)$ with an input probe $\vec{r}$ parallel to the axis $\vec{n}$ of the unitary 
$\mathsf{U}_\xi$, while in such condition the Fisher information of conventional estimation is 
expected to be zero, as addressed in Sections \ref{standqb_sec}, \ref{standqb2_sec} and 
\ref{2convqb_sec}.

\smallbreak

Figure~\ref{figFq1swi} illustrates specific conditions, when $\vec{r}\varparallel \vec{n}$, 
where the probe qubit of the switched
channel can offer an advantage over conventional estimation. Beyond, the performance of both 
approaches to estimation will vary much with the conditions, the configuration of the input probe 
$\vec{r}$, pure or mixed, in relation to the axis $\vec{n}$, the range of the parameter $\xi$, the 
level of noise, the number of passes or repetitions for conventional techniques. 
For further quantitative illustration, the performance for estimation upon measuring the probe 
qubit of the switched channel is compared below with conventional estimation upon measuring also 
a single qubit, as in Section \ref{standqb_sec} and \ref{standqb2_sec}, especially to show other 
conditions where the switched channel can bring useful contribution.

\begin{figure}[htb]
{\includegraphics[width=83mm]{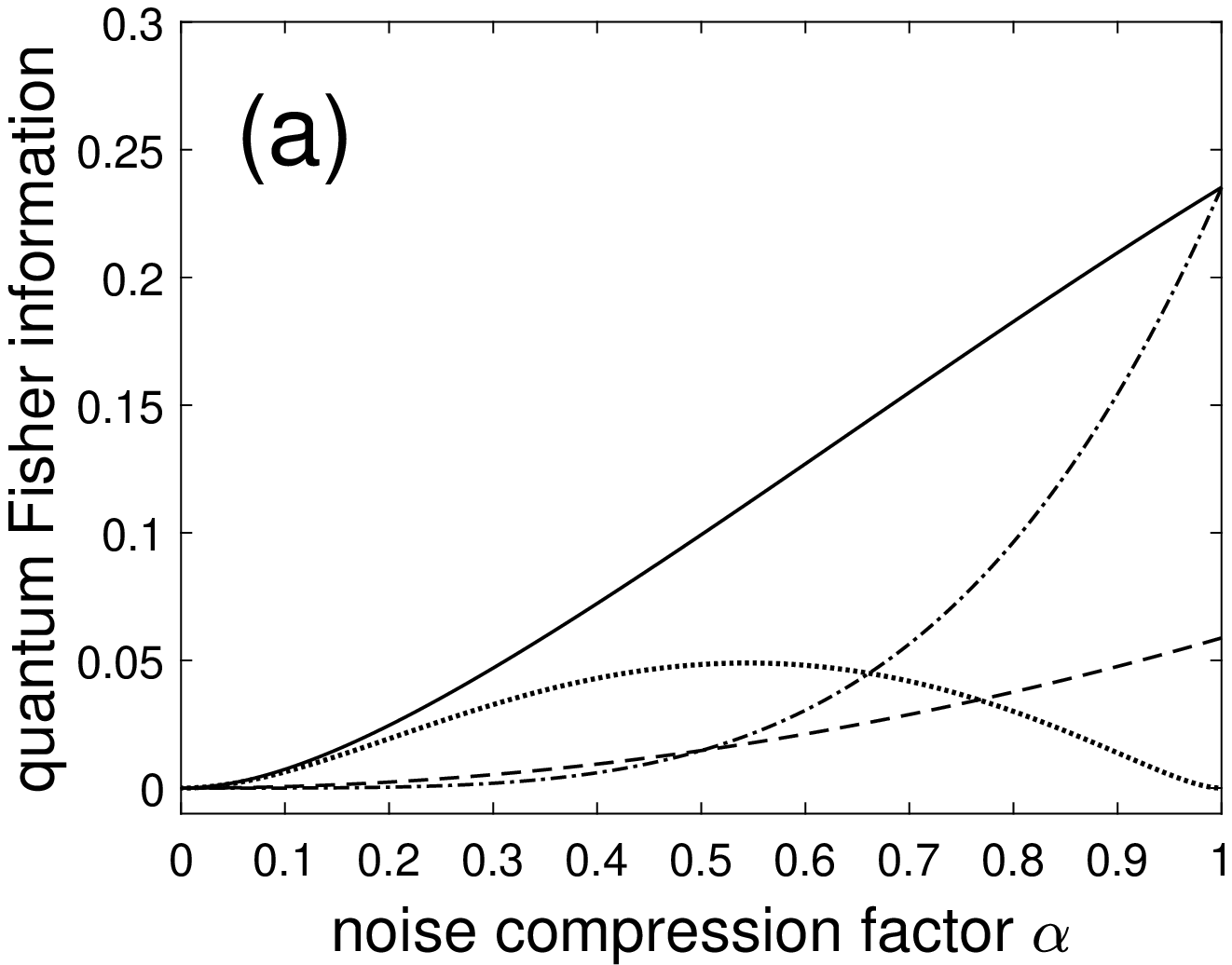}} \hfill
{\includegraphics[width=83mm]{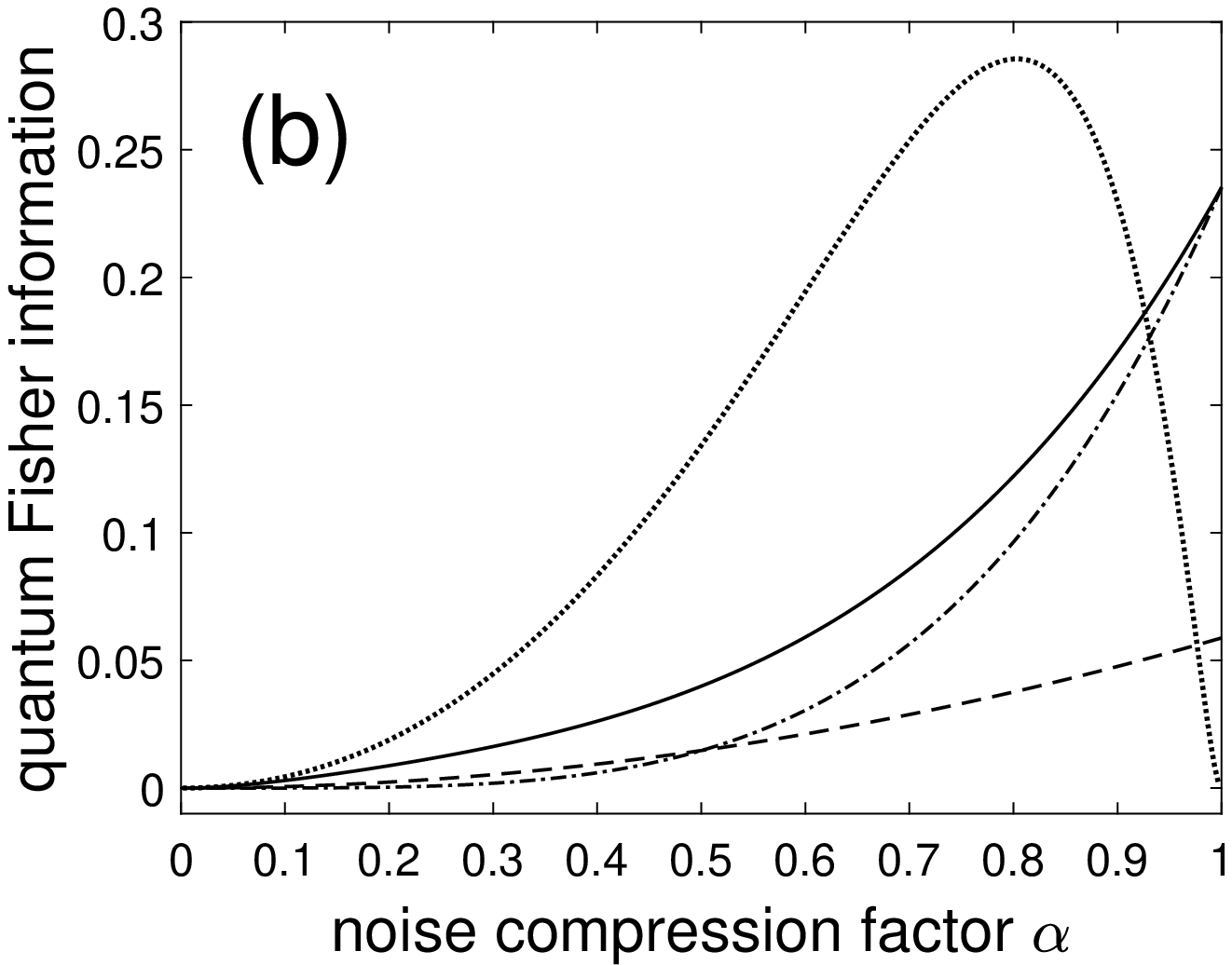}}
\caption[what appears in lof LL p177]
{As in Fig.~\ref{figFc2}, the unitary transformation $\mathsf{U}_\xi$ of Eq.~(\ref{Uxi}) is with 
axis $\vec{n}=[0.8, 0, 0.2]^\top / \sqrt{0.68}$, the input probe in Eq.~(\ref{roBloch}) is in the 
pure state with unit Bloch vector $\vec{r}=[1, 0, 0]^\top =\vec{e}_x$, and in abscissa is  
the compression factor $\alpha$ of the depolarizing noise of Eq.~(\ref{depol2}). 
For the probe qubit of the switched channel, the quantum Fisher information $F_q^{\rm swi}(\xi)$ 
from Eq.~(\ref{Fq_noisyqb}) associated with the state $\rho_+^{\rm post}$ (solid line), and 
$\rho_-^{\rm post}$ (dotted line).
From Eq.~(\ref{Fpur_qb}), the dashed line is the quantum Fisher information 
$F_q(\xi, \alpha)=\alpha^2 (\vec{n}\times\vec{r} \,)^2$
for a one-stage standard cascade as in Fig.~\ref{figUxiN}, while the dashed-dotted line
is $4F_q(2\xi, \alpha^2)=4\alpha^4 (\vec{n}\times\vec{r} \,)^2$ for a two-stage standard cascade.
The phase angle is $\xi=\pi /2$ in (a), and $\xi=\pi /4$ in (b).
}
\label{figFq1}
\end{figure}

For a unit input probe $\vec{r}$ nonparallel and nonorthogonal to the rotation axis $\vec{n}$, 
Fig.~\ref{figFq1}(a) shows that the quantum Fisher information $F_q^{\rm swi}(\xi)$ associated 
with the probe state $\rho_+^{\rm post}$ can surpass the quantum Fisher information $F_q(\xi)$ 
associated with a one-stage and a two-stage standard cascade from Fig.~\ref{figUxiN}. This 
occurs for some range of the phase $\xi$ around $\pi /2$ in Fig.~\ref{figFq1}(a), while in the 
range of $\xi$ around $\pi /4$ Fig.~\ref{figFq1}(b) shows that it is $F_q^{\rm swi}(\xi)$ of 
$\rho_-^{\rm post}$ that can dominate. 
The efficiency quantified by the Fisher information $F_q^{\rm swi}(\xi)$ in Fig.~\ref{figFq1}, of 
measuring for estimation the state $\rho_+^{\rm post}$ or the state $\rho_-^{\rm post}$, is 
determined by the geometric configuration of their respective Bloch vector 
$\vec{r}_+^{\rm \,post}$ or $\vec{r}_-^{\rm \,post}$ in $\mathbbm{R}^3$, as it
results from Eq.~(\ref{r_post}) to act in Eq.~(\ref{Fq_noisyqb}).
As observed in Fig.~\ref{figFq1}, depending on the conditions,
$\vec{r}_+^{\rm \,post}$ and $\vec{r}_-^{\rm \,post}$ change according to Eq.~(\ref{r_post}), 
and consequently $\rho_+^{\rm post}$ or $\rho_-^{\rm post}$ present more efficient 
configurations for estimation.
Also, the advantage of the switched channel observed in Fig.~\ref{figFq1} would get more 
pronounced as the input probe $\vec{r}$ tends to align with the axis $\vec{n}$.

\smallbreak

\begin{figure}[htb]
{\includegraphics[width=82mm]{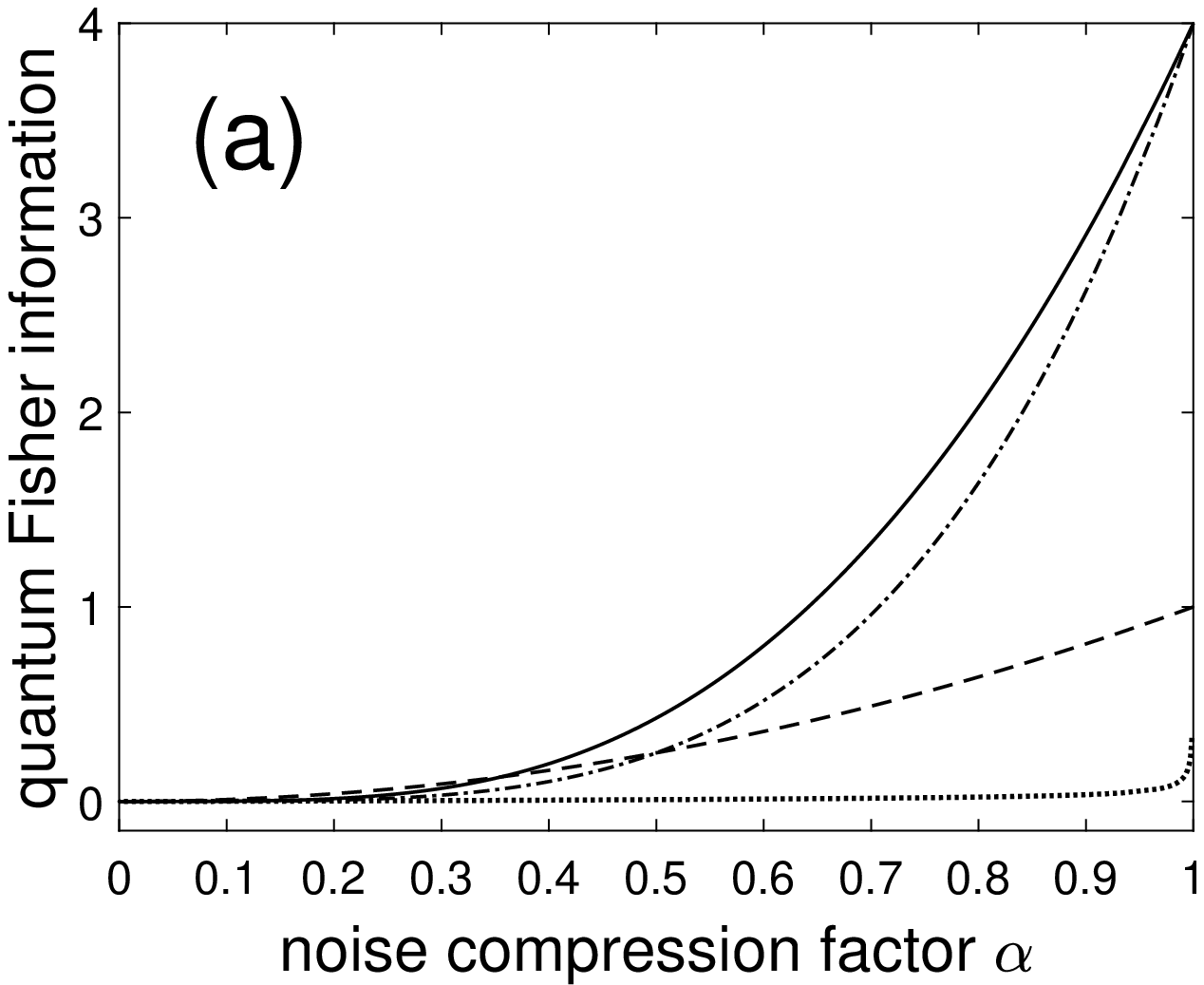}} \hfill
{\includegraphics[width=83mm]{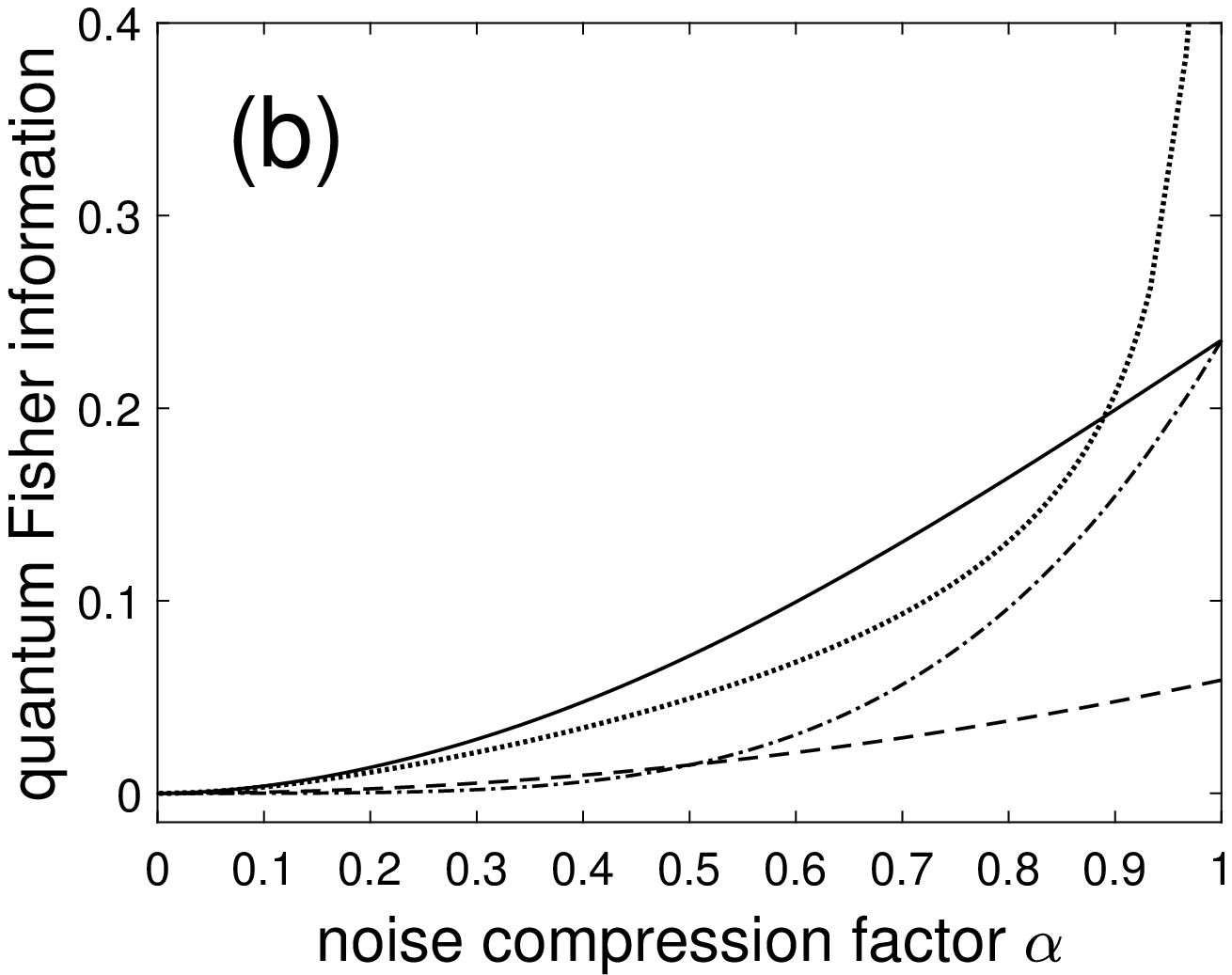}}
\caption[what appears in lof LL p177]
{The input probe in Eq.~(\ref{roBloch}) is in the pure state with unit Bloch vector 
$\vec{r}=[1, 0, 0]^\top =\vec{e}_x$, and in abscissa is the compression factor $\alpha$ of the 
depolarizing noise of Eq.~(\ref{depol2}). 
For the probe qubit of the switched channel, the quantum Fisher information $F_q^{\rm swi}(\xi)$ 
from Eq.~(\ref{Fq_noisyqb}) associated with the state $\rho_+^{\rm post}$ (solid line), and 
$\rho_-^{\rm post}$ (dotted line), after it has been averaged over the phase $\xi$ uniform in
$[0, 2\pi )$. From Eq.~(\ref{Fpur_qb}), the dashed line is the quantum Fisher information 
$F_q(\xi, \alpha)=\alpha^2 (\vec{n}\times\vec{r} \,)^2$ for a one-stage standard cascade as in 
Fig.~\ref{figUxiN}, while the dashed-dotted line is 
$4F_q(2\xi, \alpha^2)=4\alpha^4 (\vec{n}\times\vec{r} \,)^2$ for a two-stage standard cascade.
The unitary transformation $\mathsf{U}_\xi$ of Eq.~(\ref{Uxi}) is with axis 
$\vec{n}=[0, 0, 1]^\top=\vec{e}_z$ in (a), and $\vec{n}=[0.8, 0, 0.2]^\top / \sqrt{0.68}$ in (b). 
}
\label{figFq3}
\end{figure}

A comparable picture is obtained when the quantum Fisher information $F_q^{\rm swi}(\xi)$ from 
the probe state $\rho_\pm^{\rm post}$ is averaged as $\overline{F}_q\mbox{}\!^{\rm swi}$ over the 
phase $\xi$ uniform in $[0, 2\pi )$, as illustrated in Fig.~\ref{figFq3}.

\smallbreak 

In Fig.~\ref{figFq3} it can be observed that the phase-averaged quantum Fisher information 
$\overline{F}_q\mbox{}\!^{\rm swi}$ of the probe qubit from the switched channel, can still 
surpass the Fisher information from the standard cascade of Fig.~\ref{figUxiN}, either 
one-stage as $F_q(\xi, \alpha)=\alpha^2 (\vec{n}\times\vec{r} \,)^2$ or two-stage as
$4F_q(2\xi, \alpha^2)=4\alpha^4 (\vec{n}\times\vec{r} \,)^2$. This advantage is even observed 
with an optimal unit input probe $\vec{r}$ orthogonal to the axis $\vec{n}$ as shown in 
Fig.~\ref{figFq3}(a), and it gets in some respect more pronounced as $\vec{r}$ approaches 
$\vec{n}$ as illustrated in Fig.~\ref{figFq3}(b). 

When characterizing the estimation performance in the switched channel by means of the quantum 
Fisher information $F_q^{\rm swi}(\xi)$ upon measuring the probe qubit state $\rho_\pm^{\rm post}$ 
as done in Figs.~\ref{figFq1swi}--\ref{figFq3}, we have to keep in mind that $\rho_\pm^{\rm post}$ 
represent two conditional states occurring according to the probabilities $P^{\rm con}_\pm$ of 
Eq.~(\ref{Pc+a}). So the corresponding performance assessed by $F_q^{\rm swi}(\xi)$ attached to 
$\rho_+^{\rm post}$ or $\rho_-^{\rm post}$ applies also conditionally, with the probabilistic 
weights $P^{\rm con}_\pm$. At the favorable setting $p_c=1/2$ where we are, it follows from 
Eq.~(\ref{Pc+a}) that $P^{\rm con}_+$ always stays above $P^{\rm con}_-$ for any $\alpha \in [0, 1]$, 
and $P^{\rm con}_+$ goes to $1$ at low noise when $\alpha \rightarrow 1$ while $P^{\rm con}_+$ 
approaches $P^{\rm con}_-$ but stays above it at large noise when $\alpha \rightarrow 0$. So the 
post-measurement state $\rho_+^{\rm post}$ for the probe qubit is always more probable, and almost 
certain at low noise when $\alpha \rightarrow 1$. The evolutions of $F_q^{\rm swi}(\xi)$ for 
$\rho_\pm^{\rm post}$, as exemplified in Figs.~\ref{figFq1swi}--\ref{figFq3}, although they have to 
be probabilistically weighted for their interpretation, nevertheless reveal useful capabilities 
accessible by measuring the probe qubit of the switched channel for phase estimation.

For estimating the phase $\xi$ with the switched quantum channel, if one resorts to measuring the 
probe qubit as well as the control qubit, then an a-priori more efficient approach would be to 
envisage a joint measurement of the qubit pair in the entangled state
$\mathcal{S}(\rho \otimes \rho_c)$ of Eq.~(\ref{Sgenqb}). The measurement results will generally 
be governed by a $\xi$-dependent probability distribution. From there, an estimator, such as the 
maximum likelihood estimator, could be conceived for $\xi$, along with the associated Fisher 
information for an assessment of the performance. This approach is made possible based on the 
complete characterization of the joint state $\mathcal{S}(\rho \otimes \rho_c)$ worked out in 
Section~\ref{qbswitch_sec}. A joint measurement of the two qubits, control and probe, of the 
switched channel could naturally be compared with conventional estimation schemes measuring two 
qubits, active or ancilla, as in \cite{Demkowicz14} and in Section~\ref{2convqb_sec} here.
Such joint estimation involving the two qubits 
of the switched channel is however more complicated to characterize analytically, and depends on the 
choice of the joint measurement involved. We leave it as an open perspective, to come after the 
present work that has demonstrated potentialities of the switched quantum channel with indefinite 
causal order for contributing to parameter estimation and metrology.

\section{Discussion and conclusion}

In the present work we have considered a noisy unitary channel according to Fig.~\ref{figUxiN}, 
when two copies of this channel are interconnected in indefinite causal order by a quantum 
switch process as in Fig.~\ref{figSwi1}. An original contribution of the present work is, for a 
generic qubit unitary operator $\mathsf{U}_\xi$ affected by a depolarizing noise, the 
characterization of the transformation realized by the switched quantum channel of 
Fig.~\ref{figSwi1} and worked out in Section~\ref{qbswitch_sec}. This characterization lies 
essentially in the quantum operation $\mathcal{S}(\rho \otimes \rho_c)$ of Eq.~(\ref{Sgenqb}) 
acting on the joint state of the probe-control qubit pair. We have fully characterized this 
action in Bloch representation via the superoperator $\mathcal{S}_{00}(\rho)$ of 
Eq.~(\ref{NUNU_ro}), and -- for the most substantial part -- the superoperator 
$\mathcal{S}_{01}(\rho)$ of Eq.~(\ref{S01_b}) determined by $\mathcal{S}_{01}(\mathrm{I}_2)$ in 
Eq.~(\ref{S01_I2}) and $\mathcal{S}_{01}(\vec{r}\cdot \vec{\sigma})$ in Eq.~(\ref{S01_sr2}). 
This theoretical characterization of Section~\ref{qbswitch_sec} then enabled us in 
Sections~\ref{measur_sec} and \ref{Fisher_sec} to realize an analysis of the switched channel
and its performance for a task of phase estimation on the unitary operator $\mathsf{U}_\xi$ with 
depolarizing noise. A comparison has also been made with conventional techniques of estimation 
where the noisy unitary is directly probed in a one-stage or two-stage cascade with definite order, 
or several uses of them with two or more qubits.
Especially, the analysis has demonstrated three specific properties of the switched channel,
meaningful for estimation and not present with conventional techniques.

The first significant property is that the control qubit of the switched channel, although it 
never directly interacts with the unitary $\mathsf{U}_\xi$, can nevertheless be measured for the 
phase estimation on $\mathsf{U}_\xi$; and it can be measured alone, while discarding the probe 
qubit that interacts with the unitary $\mathsf{U}_\xi$ under estimation. This possibility results 
from the specific entanglement realized between the control and probe qubits by the interaction 
of the quantum switch process characterized in Section~\ref{qbswitch_sec}. This control-probe 
entangled qubit pair in the switched channel presents some similarity with conventional strategies 
as for instance analyzed in \cite{Demkowicz14}, where active probing qubits can receive assistance 
for estimation from passive ancilla qubits not directly interacting with the process under 
estimation. However, there exists an essential difference in that in such conventional estimation 
strategies, the inactive ancilla qubits must be jointly measured with the active probing qubits to
be of some use, and the ancilla qubits measured alone are inoperative for estimation. By contrast, 
in the switched channel here, the control qubit alone can be measured for estimation.

The second significant property is that the control qubit of the switched channel maintains a 
uniform efficiency for estimation, even when the input probe qubit of switched channel tends to 
align and becomes parallel with the axis $\vec{n}$ of the unitary $\mathsf{U}_\xi$. In a standard 
interaction with the qubit unitary $\mathsf{U}_\xi$ as in Eq.~(\ref{r1}), rotation of a probe 
qubit $\vec{r}$ parallel to the rotation axis $\vec{n}$ issues no physical effect or information 
enabling any access to the rotation angle $\xi$. This mechanism of rotating an isolated qubit does 
not hold with the switched quantum channel, which rather acts on a probe-control entangled qubit 
pair in the specific way characterized in Section~\ref{qbswitch_sec} enabling estimation even with 
an input probe with $\vec{r}\varparallel \vec{n}$.

The third significant property is that the control qubit of the switched channel also maintains a 
uniform efficiency for estimation, even when the input probe qubit of switched channel tends to 
depolarize or even becomes completely depolarized when $\| \vec{r}\, \|=0$. Conventional techniques 
based on the qubit interaction $\rho \mapsto \mathsf{U}_\xi \rho \mathsf{U}_\xi^\dagger$ have 
no effect on the completely depolarized input probe $\rho=\mathrm{I}_2 /2$, which remains invariant 
and insensitive to the phase $\xi$ and thus inoperative for estimating $\xi$.
By contrast, the specific interaction in the switched channel with indefinite order we
characterized in Section~\ref{qbswitch_sec} makes it possible to use a fully depolarized input 
probe for estimation.

Conventional estimation techniques, as addressed in Sections \ref{standqb_sec}, \ref{standqb2_sec} 
and \ref{2convqb_sec}, although useful in their own right, do not share these properties of the 
switched channel, and they become gradually inoperative when the input probe tends to align with 
the unitary $\mathsf{U}_\xi$ under estimation or progressively depolarizes.
These three properties relevant to estimation are essentially contributed by the control qubit of 
the switched channel, and they manifest specific and unusual capabilities for quantum information
processing observable in channels with coherent control of indefinite causal order, as also 
observed for instance in \cite{Ebler18}. We have also analyzed in Section~\ref{FiProbe_sec} the 
measurement of the probe qubit of the switched channel and showed it can add useful capabilities for 
phase estimation. 
These specific and unusual properties of the switched channel for estimation, and specially the 
capabilities of its control qubit, stem from the specific interaction entangling the two qubits
in the joint state $\mathcal{S}(\rho \otimes \rho_c)$ of Eq.~(\ref{Sgenqb}) 
characterized in Section~\ref{qbswitch_sec}. An essential ingredient is the qubit superoperator 
$\mathcal{S}_{01}(\cdot)$ of Eqs.~(\ref{S01}) and (\ref{S01_b}), which is a specific interaction 
term following from the coherent superposition of causal orders implemented by the switch process 
according to Eqs.~(\ref{Wjk})--(\ref{Sgen1}). 
This interaction term $\mathcal{S}_{01}(\cdot)$ acts in the joint state
$\mathcal{S}(\rho \otimes \rho_c)$ of Eq.~(\ref{Sgenqb}) only when there exists 
an effective coherent superposition of orders, at $p_c \not = 0, 1$ for the control qubit.
The switch process of Eqs.~(\ref{Wjk})--(\ref{Sgen1}) then mixes two elementary channels having 
Kraus operators that do not commute, as explained at the end of Section~\ref{qbswitch_sec}.
Their combinations in $\mathcal{S}_{01}(\cdot)$ of Eqs.~(\ref{S01}) and (\ref{S01_b}) superpose 
various paths across the operators determining the transmission by the composite switched channel. 
In particular, these combinations of paths in Eqs.~(\ref{S01}) and (\ref{S01_b}) lead for the 
interaction term to $\mathcal{S}_{01}(\mathrm{I}_2) \not = \mathrm{I}_2$, as expressed by 
Eq.~(\ref{S01_I2}), which has a direct impact for the transmission of the fully depolarized input 
probe $\rho=\mathrm{I}_2 /2$, and also more generally of the generic input probe $\rho$ of 
Eq.~(\ref{roBloch}).
With the generic input probe of Eq.~(\ref{roBloch}), this specific interaction in the switched 
channel via the superoperator $\mathcal{S}_{01}(\rho)$ of Eq.~(\ref{S01_b}), leads to a control 
qubit in the state $\rho^{\rm con}_\xi$ of Eqs.~(\ref{Sgenqb_tp1})--(\ref{Sgenqb_tp2}), 
which is unaffected by $\vec{r}$ but sensitive to $\xi$, essentially by way of 
$\tr[\mathcal{S}_{01}(\rho)]=\tr[\mathcal{S}_{01}(\mathrm{I}_2)/2]=Q_\xi(\alpha) \not = 1$, 
for any $\vec{r}$.
As a result, measurement of the control qubit in the state $\rho^{\rm con}_\xi$, 
as governed by the probabilities $P^{\rm con}_\pm$ of Eqs.~(\ref{Pc+})--(\ref{Pc+a}), is 
unaffected by $\vec{r}$, but remains sensitive to $\xi$, even when $\vec{r}\varparallel \vec{n}$ 
and $\| \vec{r}\, \|=0$, affording estimation capabilities in these conditions.

\bigbreak

The characterization of $\mathcal{S}(\rho \otimes \rho_c)$ carried out in 
Section~\ref{qbswitch_sec} with the important reference formed by the depolarizing noise, 
could be extended to other qubit noise models. Arbitrary Pauli noises could be incorporated
with three distinct probabilities $p_x$, $p_y$, $p_z$ for the three Pauli operators in
Eq.~(\ref{depol1}) instead of the common probability $p/3$. Then the derivation of 
Section~\ref{qbswitch_sec} could proceed in a similar way, except at the stage where relations 
like Eqs.~(\ref{WxWx+}) and (\ref{WxyWxy+}) would have to be incorporated into Eq.~(\ref{S01_b}) 
in accordance with the specific probabilistic weights $p_x$, $p_y$ or $p_z$; and similarly for 
Eq.~(\ref{Wl_Wl_rs}) and Eqs.~(\ref{Wxy+})--(\ref{Wzx}). This would lead to more bulky expressions,
but useful capabilities of the switched channel can be expected to be preserved. We have 
explicitly tested a bit-flip noise and a phase-flip noise, and verified that both preserve the 
essential capabilities of the switched channel with indefinite causal order useful to estimation, 
yet with added variability in the detailed performance especially depending on the angle between 
the privileged axis of the noise and the orientation $\vec{n}$ of the unitary $\mathsf{U}_\xi$.

This robustness with various noise models is also an interesting feature to robustly preserve the 
capabilities for estimation from the switched channel in concrete physical implementations. Also, 
as observed in Section~\ref{Fisher_sec}, a defective preparation (pure or mixed) of the input probe 
qubit will have no effect on the efficiency for estimation from the control qubit. In addition, as 
observed in Section~\ref{measur_sec}, an indefinite superposition of causal orders robustly takes 
place for any $p_c \in (0, 1)$ in the preparation of the control qubit 
$\ket{\psi_c}=\sqrt{p_c}\ket{0}+\sqrt{1-p_c}\ket{1}$. From these properties, it can be 
expected that the capabilities useful to estimation of the switched channel will be robustly 
preserved in physical implementations with experimental imperfections.

\bigbreak

Previous works on quantum switched channels with indefinite order concentrated first on 
communication of information \cite{Ebler18,Procopio19,Procopio20,Loizeau20}, and more recently on 
parameter estimation yet in significantly distinct processes and conditions 
\cite{Frey19,Mukhopadhyay18,Zhao20}. By comparison, specificities of our study are that it 
considers a qubit system, experiencing a generic qubit unitary operator $\mathsf{U}_\xi$, 
characterized by an unknown parameter $\xi$ to be estimated. Parameter estimation on a qubit 
unitary process represents an important reference task for quantum metrology, which is investigated 
here for the first time in quantum switched channels with indefinite order. In addition, in the 
estimation task, the realistic condition where noise is present is taken into account as a 
significant specificity.

The capability of the switched channel for estimation with a fully depolarized input probe is 
reminiscent of the situation of the quantum communication channel analyzed for instance in 
\cite{Ebler18} as evoked in the Introduction. In \cite{Ebler18}, an isolated communication channel 
is by itself fully depolarizing and unable to transmit any useful information, with a zero Holevo 
information or information capacity. When duplicated and inserted in a quantum switch process as 
in Fig.~\ref{figSwi1} with indefinite causal order, it gives rise to a quantum channel with nonzero 
information capacity, enabling effective transmission of information. 
We observe a comparable behavior here, for an estimation task rather than a communication task, 
assessed by the Fisher information instead of the information capacity. 
The effect in \cite{Ebler18} is observed with a coherent control to superpose two depolarizing 
channels in indefinite causal order. In a recent study \cite{Abbott20} related to \cite{Ebler18}, a 
comparable effect is observed with no indefinite causal order, but instead with a coherent control 
to determine which of the two depolarizing channels is traversed. These are two distinct phenomena, 
which can be observed separately as in \cite{Ebler18} and in \cite{Abbott20}, and which are further 
discussed in \cite{Kristjansson20,Loizeau20}. The effects for estimation in our study occur in 
the presence of 
coherent control of indefinite causal order, as in \cite{Ebler18}. Examining if the effects we 
observe would occur or not in the setting of \cite{Abbott20} having coherent control but no 
indefinite causal order, could possibly help to disentangle the respective roles of coherent control 
and indefinite order. Such an extension would contribute to the exploration of the properties and 
capabilities of coherent quantum superposition of processing channels, with the structure of either 
\cite{Ebler18} or \cite{Abbott20}.

\bigbreak
The specific capabilities uncovered here for estimation will be practically accessible at the cost 
of physically implementing the quantum switch process. In this respect, techniques have been 
proposed that are particularly appealing for photonic implementations, as in \cite{Procopio15} for 
instance, with an interferometric setup, by combining a spatial mode of a photon for the control 
with its polarization for the probe. Employing such setups for estimation could render practically 
accessible the useful properties of the switched process in photon metrology and interferometry, 
where qubit phase estimation is an essential operation, at the root of many applications such as 
high-sensitivity and high-precision measurements, atomic clocks, frequency standards 
\cite{Giovannetti06,Giovannetti11,DAriano98,vanDam07,Chapeau15,Degen17}.
In particular, if the probe and control qubits can physically be made sufficiently distant when 
the probe interacts with the unitary, specially interesting properties could follow.

In this way, the present study contributes to the identification and analysis of the properties 
and capabilities of switched quantum channels with indefinite causal order for quantum signal 
and information processing, along with new possibilities useful to quantum estimation and qubit 
metrology.

\section*{Appendix A}

\renewcommand{\theequation}{A-\arabic{equation}} 
\setcounter{equation}{0}  

In this Appendix we work out the Bloch representation for $\mathcal{S}_{01}(\mathrm{I}_2)$ and 
$\mathcal{S}_{01}(\vec{r}\cdot \vec{\sigma} )$ characterizing the superoperator 
$\mathcal{S}_{01}(\rho)$ of Eq.~(\ref{S01_b}).

\bigbreak \noindent  
$\bullet$ For $\mathcal{S}_{01}(\mathrm{I}_2)\,$:

By unitarity of $\mathsf{U}_\xi$, in Eq.~(\ref{W0}) one has 
$\mathcal{W}_0(\mathrm{I}_2)=\mathrm{I}_2$. In Eq.~(\ref{Wl}) one has 
$\mathcal{W}_\ell(\mathrm{I}_2)=\sigma_\ell \mathsf{U}_\xi\sigma_\ell\mathsf{U}_\xi^\dagger$.
Terms comparable to $\sigma_\ell \mathsf{U}_\xi$ can be evaluated via Eq.~(\ref{Uxi}) and the 
standard behavior of products of Pauli operators. One obtains, for $\mathcal{W}_x(\mathrm{I}_2)$ 
for instance,
\begin{equation}
\sigma_x \mathsf{U}_\xi = \cos(\xi/2) \sigma_x + \sin(\xi/2)
(-n_z \sigma_y +n_y \sigma_z -i n_x \mathrm{I}_2 ) \;,
\label{sxU}
\end{equation}
and also
\begin{equation}
\sigma_x \mathsf{U}_\xi^\dagger = \cos(\xi/2) \sigma_x + \sin(\xi/2)
(n_z \sigma_y -n_y \sigma_z +i n_x \mathrm{I}_2 ) \;,
\label{sxU+}
\end{equation}
which gives
\begin{equation}
\sigma_x \mathsf{U}_\xi \sigma_x \mathsf{U}_\xi^\dagger = 
\left(1-2\sin^2\Bigl(\frac{\xi}{2} \Bigr)(1-n_x^2) \right)\mathrm{I}_2 +
i2\sin\Bigl(\frac{\xi}{2} \Bigr)\left( 
\cos\Bigl(\frac{\xi}{2} \Bigr) \begin{bmatrix} 0 \\ n_y \\ n_z \end{bmatrix} + 
\sin\Bigl(\frac{\xi}{2} \Bigr) n_x \begin{bmatrix} 0 \\ -n_z \\ n_y \end{bmatrix} 
\right) \cdot \vec{\sigma} \;,
\label{sxUsxU+2}
\end{equation}
resulting in
\begin{equation}
\bigl(\mathcal{W}_x+\mathcal{W}_x^\dagger \bigr)(\mathrm{I}_2)=
2\bigl[1-2\sin^2(\xi/2)(1-n_x^2) \bigr]\mathrm{I}_2 \;.
\label{WxWx+}
\end{equation}
Similar relations can be obtained for $\mathcal{W}_y(\mathrm{I}_2)$ and 
$\mathcal{W}_z(\mathrm{I}_2)$ and their adjoints, and due to the isotropic action of the
depolarizing noise in $\mathbbm{R}^3$, they add up uniformly in Eq.~(\ref{S01_b}) so as
to give
\begin{equation}
\sum_{\ell=x,y,z} \bigl(\mathcal{W}_\ell + \mathcal{W}_\ell^\dagger \bigr)(\mathrm{I}_2) =
2\bigl[3-4\sin^2(\xi/2) \bigr]\mathrm{I}_2 \;.
\label{S_WxWx+}
\end{equation}

For $\mathcal{W}_{\ell\ell'}(\mathrm{I}_2)$ in Eq.~(\ref{Wll}) one has for instance 
$\mathcal{W}_{xy}(\mathrm{I}_2)= \sigma_x \mathsf{U}_\xi \sigma_y \sigma_x \mathsf{U}_\xi^\dagger 
\sigma_y = -i\sigma_x \mathsf{U}_\xi \sigma_z \mathsf{U}_\xi^\dagger \sigma_y $.
Via the circular permutation behavior among Pauli operators and their products,
relations analogous to Eqs.~(\ref{sxU})--(\ref{sxU+}) readily follow, as for instance
\begin{equation}
\sigma_z \mathsf{U}_\xi^\dagger = \cos(\xi/2) \sigma_z + \sin(\xi/2)
(n_y \sigma_x -n_x \sigma_y +i n_z \mathrm{I}_2 ) \;.
\label{szU+}
\end{equation}
From Eqs.~(\ref{sxU}) and (\ref{szU+}) one has
\begin{eqnarray}
\nonumber
\sigma_x \mathsf{U}_\xi \sigma_z \mathsf{U}_\xi^\dagger &=& 
-i\cos^2(\xi/2) \sigma_y + 2\cos(\xi/2)\sin(\xi/2) (n_y \mathrm{I}_2 -in_x \sigma_z) \\
\label{sxUszU+}
\mbox{} &+& \sin^2(\xi/2) \bigl[2n_x n_z \mathrm{I}_2 +i(1-2n_z^2)\sigma_y+i2n_y n_z \sigma_z
\bigr] \;,
\end{eqnarray}
and by right-multiplying by $-i\sigma_y$ one obtains
\begin{eqnarray}
\nonumber
\mathcal{W}_{xy}(\mathrm{I}_2) &=& 
\bigl[-\cos^2(\xi/2) +\sin^2(\xi/2)(1-2n_z^2)\bigr] \mathrm{I}_2 \\
\label{Wxy}
\mbox{} &+& i2\bigl[ \cos(\xi/2)\sin(\xi/2) (n_x \sigma_x-n_y\sigma_y) -\sin^2(\xi/2)
n_z (n_y \sigma_x +n_x \sigma_y ) \bigr] \;,
\end{eqnarray}
yielding
\begin{equation}
\bigl(\mathcal{W}_{xy}+\mathcal{W}_{xy}^\dagger \bigr)(\mathrm{I}_2)= 
2\bigl[-1+2\sin^2(\xi/2)(1-n_z^2) \bigr]\mathrm{I}_2 \;.
\label{WxyWxy+}
\end{equation}
Two other relations similar to Eq.~(\ref{WxyWxy+}) can be obtained, which again, due to the 
isotropic action of the depolarizing noise in $\mathbbm{R}^3$, add up uniformly in 
Eq.~(\ref{S01_b}) to provide
\begin{equation}
\bigl[
\bigl(\mathcal{W}_{xy} + \mathcal{W}_{xy}^\dagger \bigr) +
\bigl(\mathcal{W}_{yz} + \mathcal{W}_{yz}^\dagger \bigr) +
\bigl(\mathcal{W}_{zx} + \mathcal{W}_{zx}^\dagger \bigr) \bigr](\mathrm{I}_2) =
2\bigl[-3+4\sin^2(\xi/2) \bigr]\mathrm{I}_2 \;.
\label{S_WxWy+}
\end{equation}
Meanwhile $\mathcal{W}_{\ell\ell}(\mathrm{I}_2)=\mathrm{I}_2$ for each $\ell =x, y, z$. Finally, 
by gathering the pieces, in Eq.~(\ref{S01_b}) one obtains
\begin{eqnarray}
\label{S01_I1}
\mathcal{S}_{01}(\mathrm{I}_2) &=& 
(1-p)^2 \mathrm{I}_2 + (1-p)\dfrac{p}{3} 2\bigl[3-4\sin^2(\xi/2) \bigr]\mathrm{I}_2
+ \Bigl(\dfrac{p}{3} \Bigr)^2 \bigl[-3+8\sin^2(\xi/2)\bigr] \mathrm{I}_2 \\
\label{S01_I2_A}
&=& \biggl[ \dfrac{4}{3}p\Bigl(1-\dfrac{4}{3}p \Bigr)\cos(\xi)
+1-\dfrac{4}{3}p\Bigl(1-\dfrac{p}{3} \Bigr) \biggr] \mathrm{I}_2 \;,
\end{eqnarray}
with Eq.~(\ref{S01_I2_A}) which is returned to the main text as Eq.~(\ref{S01_I2}).

\bigbreak \noindent  
$\bullet$ For $\mathcal{S}_{01}(\vec{r}\cdot \vec{\sigma})$\,:

In  Eq.~(\ref{W0}) one has 
$\mathcal{W}_0(\vec{r}\cdot \vec{\sigma})=U_\xi^2\vec{r}\cdot \vec{\sigma}$. 
In Eq.~(\ref{Wl}) one has 
$\mathcal{W}_\ell(\vec{r}\cdot \vec{\sigma})=\sigma_\ell \mathsf{U}_\xi 
(U_\xi \vec{r}\cdot \vec{\sigma})\sigma_\ell\mathsf{U}_\xi^\dagger$.
To further handle such equation, it is useful to characterize, for a generic Bloch vector
$\vec{r} \in \mathbbm{R}^3$, operators of the form $\sigma_\ell(\vec{r}\cdot \vec{\sigma})$
or $(\vec{r}\cdot \vec{\sigma})\sigma_\ell$. For instance, one has
\begin{equation}
\sigma_x(\vec{r}\cdot \vec{\sigma}) =r_x\mathrm{I}_2+ir_y \sigma_z-ir_z \sigma_y \;.
\label{sx_rs}
\end{equation}
Such relation can usefully be written in matrix notation as
\begin{equation}
\sigma_x(\vec{r}\cdot \vec{\sigma}) =[\vec{r}\,]_x \mathrm{I}_2+iS_x\vec{r}\cdot \vec{\sigma}\;,
\label{sx_rs_}
\end{equation}
with the $3\times 3$ real matrix $S_x$ defined in Eq.~(\ref{matSx}) of Appendix~B, and 
$[\cdot]_x$ which represents the $x$ component of a vector in $\mathbbm{R}^3$. The matrix 
notation makes transparent the way such relation as Eq.~(\ref{sx_rs_}) extends to 
$\sigma_\ell(\vec{r}\cdot \vec{\sigma})$, or to $(\vec{r}\cdot \vec{\sigma})\sigma_\ell$ by
conjugate transposition, for $\ell =x, y, z$. In this way, one has the chain transformations
\begin{eqnarray}
\label{Wl_rs1}
\mathcal{W}_\ell(\vec{r}\cdot \vec{\sigma}) &=&
\sigma_\ell \mathsf{U}_\xi (U_\xi \vec{r}\cdot \vec{\sigma})\sigma_\ell\mathsf{U}_\xi^\dagger \\
\label{Wl_rs2}
&=& \sigma_\ell \mathsf{U}_\xi\bigl( 
[U_\xi \vec{r}\,]_\ell \mathrm{I}_2 -iS_\ell U_\xi \vec{r}\cdot \vec{\sigma}\, \bigr) 
\mathsf{U}_\xi^\dagger \\
\label{Wl_rs3}
&=& \sigma_\ell \bigl( 
[U_\xi \vec{r}\,]_\ell \mathrm{I}_2 -iU_\xi S_\ell U_\xi \vec{r}\cdot \vec{\sigma}\, \bigr) \\
\label{Wl_rs4}
&=& [U_\xi \vec{r}\,]_\ell \sigma_\ell -i[U_\xi S_\ell U_\xi \vec{r}\,]_\ell \mathrm{I}_2 +
S_\ell U_\xi S_\ell U_\xi \vec{r}\cdot \vec{\sigma} \;.
\end{eqnarray}

For Eq.~(\ref{S01_b}) one therefore obtains
\begin{equation}
\bigl(\mathcal{W}_\ell+\mathcal{W}_\ell^\dagger \bigr)(\vec{r}\cdot \vec{\sigma}) =
2(\eM_\ell U_\xi + S_\ell U_\xi S_\ell U_\xi ) \vec{r}\cdot \vec{\sigma} \;,
\label{Wl_Wl_rs}
\end{equation}
with the three $3\times 3$ projection matrices $\eM_\ell$ having as entries $0$ everywhere except 
a single $1$ at row $\ell$ and column $\ell$, for $\ell=x, y, z$. 
By the isotropy of the depolarizing noise in $\mathbbm{R}^3$, the terms similar to
Eq.~(\ref{Wl_Wl_rs}), for $\ell=x, y, z$, add up uniformly in Eq.~(\ref{S01_b}). And since 
$\eM_x+\eM_y+\eM_z=I_3$ the $3\times 3$ identity matrix on $\mathbbm{R}^3$, this leads to
\begin{equation}
\sum_{\ell=x,y,z}\bigl(\mathcal{W}_\ell+\mathcal{W}_\ell^\dagger\bigr)(\vec{r}\cdot \vec{\sigma}) =
2\biggl(I_3 +\sum_{\ell=x,y,z}S_\ell U_\xi S_\ell \biggl)U_\xi\vec{r}\cdot \vec{\sigma} \;.
\label{S_Wl_Wl_rs}
\end{equation}

We now turn in Eq.~(\ref{Wll}) to $\mathcal{W}_{\ell \ell'}(\vec{r}\cdot \vec{\sigma})=
\sigma_\ell \mathsf{U}_\xi \sigma_{\ell'} \mathsf{U}_\xi (\vec{r}\cdot \vec{\sigma}) 
\mathsf{U}_\xi^\dagger \sigma_\ell \mathsf{U}_\xi^\dagger \sigma_{\ell'}$. Here, it is useful to 
characterize, for a generic Bloch vector $\vec{r} \in \mathbbm{R}^3$, operators of the form 
$\sigma_\ell(\vec{r}\cdot \vec{\sigma})\sigma_{\ell'}$ for $\ell \not = \ell'$. For instance, 
from Eq.~(\ref{sx_rs_}) one further obtains
\begin{eqnarray}
\label{sx_rs_sy_1}
\sigma_x(\vec{r}\cdot \vec{\sigma})\sigma_y &=& 
[\vec{r}\,]_x \sigma_y +i[S_x\vec{r}\,]_y \mathrm{I}_2 +S_y S_x\vec{r}\cdot \vec{\sigma}\;,\\
\label{sx_rs_sy_2}
&=& (\eM_{yx}+S_y S_x)\vec{r}\cdot \vec{\sigma} +i[S_x\vec{r}\,]_y \mathrm{I}_2 \;,
\end{eqnarray}
where we have defined $[\vec{r}\,]_\ell \sigma_{\ell'} = \eM_{\ell'\ell} \vec{r}\cdot \vec{\sigma}$ 
for a generic Bloch vector $\vec{r} \in \mathbbm{R}^3$, with the $3\times 3$ real matrix 
$\eM_{\ell'\ell}$ having as entries $0$ everywhere except a single $1$ at row $\ell'$ and column 
$\ell$ (and $\eM_{\ell\ell}$ is like $\eM_\ell$ above).  
The three matrices $S_\ell$ defined by Eqs.~(\ref{matSx})--(\ref{matSz}) of Appendix~B
satisfy $S_\ell S_{\ell'}=\eM_{\ell'\ell}$ for any $\ell \not = \ell'$. We also define the
$3\times 3$ real symmetric matrix $T_{\ell\ell'}=\eM_{\ell\ell'}+\eM_{\ell'\ell}$ for any
$\ell, \ell' =x, y, z$. As an extension to Eq.~(\ref{sx_rs_sy_2}) follows for $\ell \not = \ell'$
the generic relation
\begin{eqnarray}
\sigma_\ell(\vec{r}\cdot \vec{\sigma})\sigma_{\ell'} =
T_{\ell\ell'}\vec{r}\cdot \vec{\sigma} +i[S_\ell\vec{r}\,]_{\ell'} \mathrm{I}_2 \;.
\label{sl_rs_sl'}
\end{eqnarray}

One now has the chain transformations  
\begin{eqnarray}
\label{Wll_rs1}
\mathcal{W}_{\ell \ell'}(\vec{r}\cdot \vec{\sigma}) &=& 
\sigma_\ell \mathsf{U}_\xi \sigma_{\ell'} 
(U_\xi \vec{r}\cdot \vec{\sigma})
\sigma_\ell \mathsf{U}_\xi^\dagger \sigma_{\ell'} \\
\label{Wll_rs2}
&=& \sigma_\ell \mathsf{U}_\xi \bigl(
T_{\ell\ell'} U_\xi \vec{r}\cdot \vec{\sigma} +i[S_{\ell'}U_\xi\vec{r}\,]_\ell \mathrm{I}_2
\bigr) \mathsf{U}_\xi^\dagger \sigma_{\ell'} \\
\label{Wll_rs3}
&=& \sigma_\ell \bigl(U_\xi 
T_{\ell\ell'} U_\xi \vec{r}\cdot \vec{\sigma} +i[S_{\ell'}U_\xi\vec{r}\,]_\ell \mathrm{I}_2
\bigr) \sigma_{\ell'} \\
\label{Wll_rs4}
&=& T_{\ell\ell'} U_\xi T_{\ell\ell'} U_\xi \vec{r}\cdot \vec{\sigma} 
+i[S_\ell U_\xi T_{\ell\ell'} U_\xi \vec{r}\,]_{\ell'} \mathrm{I}_2
+i[S_{\ell'}U_\xi\vec{r}\,]_\ell \sigma_\ell \sigma_{\ell'} \;.
\end{eqnarray}
The last term $\displaystyle i[S_{\ell'}U_\xi\vec{r}\,]_\ell \sigma_\ell \sigma_{\ell'}$ in 
Eq.~(\ref{Wll_rs4}) evaluates as follows. When $\ell\ell' =xy$ it is 
$-[S_yU_\xi\vec{r}\,]_x \sigma_z=-\eM_{zx}S_y U_\xi\vec{r}\cdot \vec{\sigma}
=-\eM_z U_\xi\vec{r}\cdot \vec{\sigma}$, in a similar way when $\ell\ell' =yz$ it is 
$-\eM_x U_\xi\vec{r}\cdot \vec{\sigma}$, and when $\ell\ell' =zx$ it is 
$-\eM_y U_\xi\vec{r}\cdot \vec{\sigma}$.

For Eq.~(\ref{S01_b}) this leads to
\begin{eqnarray}
\label{Wxy+}
\bigl(\mathcal{W}_{xy}+\mathcal{W}_{xy}^\dagger\bigr)(\vec{r}\cdot \vec{\sigma}) &=&
2(T_{xy} U_\xi T_{xy}-\eM_z) U_\xi \vec{r}\cdot \vec{\sigma} \;, \\
\label{Wyz+}
\bigl(\mathcal{W}_{yz}+\mathcal{W}_{yz}^\dagger\bigr)(\vec{r}\cdot \vec{\sigma}) &=&
2(T_{yz} U_\xi T_{yz}-\eM_x) U_\xi \vec{r}\cdot \vec{\sigma} \;, \\
\label{Wzx}
\bigl(\mathcal{W}_{zx}+\mathcal{W}_{zx}^\dagger\bigr)(\vec{r}\cdot \vec{\sigma}) &=&
2(T_{zx} U_\xi T_{zx}-\eM_y) U_\xi \vec{r}\cdot \vec{\sigma} \;,
\end{eqnarray}
these three terms summing uniformly by the isotropy of the depolarizing noise, to give
\begin{equation}
2(T_{xy}U_\xi T_{xy}+T_{yz}U_\xi T_{yz}+T_{zx}U_\xi T_{zx}-I_3)U_\xi \vec{r}\cdot \vec{\sigma} \;.
\label{S_Wxy+}
\end{equation}

For Eq.~(\ref{S01_b}) finally, for the three terms 
$\mathcal{W}_{\ell\ell}(\vec{r}\cdot \vec{\sigma})$ resulting from Eq.~(\ref{Wll}), one has 
$\mathcal{W}_{xx}(\vec{r}\cdot \vec{\sigma})=(XU_\xi)^2 \vec{r}\cdot \vec{\sigma}$, also
$\mathcal{W}_{yy}(\vec{r}\cdot \vec{\sigma})=(YU_\xi)^2 \vec{r}\cdot \vec{\sigma}$ and
$\mathcal{W}_{zz}(\vec{r}\cdot \vec{\sigma})=(ZU_\xi)^2 \vec{r}\cdot \vec{\sigma}$, with the 
three $3\times 3$ real matrices $X$, $Y$ and $Z$ defined by Eqs.~(\ref{matX})--(\ref{matZ}) of 
Appendix~B.

By gathering the pieces, one obtains with Eq.~(\ref{S01_b}),
\begin{eqnarray}
\nonumber
\mathcal{S}_{01}(\vec{r}\cdot \vec{\sigma}) &=& 
(1-p)^2 U_\xi^2 \vec{r}\cdot \vec{\sigma} + (1-p)\dfrac{p}{3} 
2\biggl(I_3 +\sum_{\ell=x,y,z}S_\ell U_\xi S_\ell \biggl)U_\xi\vec{r}\cdot \vec{\sigma} \\
\nonumber
&+& \Bigl(\dfrac{p}{3} \Bigr)^2 \bigl[
2(T_{xy}U_\xi T_{xy}+T_{yz}U_\xi T_{yz}+T_{zx}U_\xi T_{zx}-I_3)U_\xi \vec{r}\cdot \vec{\sigma} \\
\label{S01_sr1}
&+& 
(XU_\xi X+YU_\xi Y+ZU_\xi Z)U_\xi \vec{r}\cdot \vec{\sigma} \bigr] \;,
\end{eqnarray}
which can also be written as
\begin{equation}
\mathcal{S}_{01}(\vec{r}\cdot \vec{\sigma}) = \Bigl[
(1-p)^2 U_\xi + (1-p)\dfrac{p}{3} 2 \bigl[I_3+L_1(U_\xi) \bigr]+
\Bigl(\dfrac{p}{3} \Bigr)^2 \bigl[ 2L_2(U_\xi)-2I_3 +L_3(U_\xi) \bigr]
\Bigr] U_\xi \vec{r}\cdot \vec{\sigma} \;,
\label{S01_sr2_A}
\end{equation}
where we have defined the three fixed linear transformations of the $3\times 3$ matrix
$U_\xi$ as
\begin{eqnarray}
\label{L1U}
L_1(U_\xi) &=& S_x U_\xi S_x + S_y U_\xi S_y +S_z U_\xi S_z \;, \\
\label{L2U}
L_2(U_\xi) &=&
T_{xy}U_\xi T_{xy}+T_{yz}U_\xi T_{yz}+T_{zx}U_\xi T_{zx} \;, \\
\label{L3U}
L_3(U_\xi) &=& XU_\xi X+YU_\xi Y+ZU_\xi Z \;.
\end{eqnarray}
The final characterization of Eq.~(\ref{S01_sr2_A}) is returned to the main text as
Eq.~(\ref{S01_sr2}).

\section*{Appendix B}

\renewcommand{\theequation}{B-\arabic{equation}} 
\setcounter{equation}{0}  

We define the $3\times 3$ real matrices
\begin{eqnarray}
\label{matSx}
S_x =
\begin{bmatrix}
0 & 0 & 0 \\
0 & 0 & -1 \\
0 & 1 & 0
\end{bmatrix} \;, \\
\label{matSy}
S_y =
\begin{bmatrix}
0 & 0 & 1 \\
0 & 0 & 0 \\
-1 & 0 & 0
\end{bmatrix} \;, \\
\label{matSz}
S_z =
\begin{bmatrix}
0 & -1 & 0 \\
1 & 0 & 0 \\
0 & 0 & 0
\end{bmatrix} \;,
\end{eqnarray}
and the three diagonal matrices
\begin{eqnarray}
\label{matX}
X=
\begin{bmatrix}
1 & 0 & 0 \\
0 & -1 & 0 \\
0 & 0 & -1
\end{bmatrix} \;, \\
\label{matY}
Y=
\begin{bmatrix}
-1 & 0 & 0 \\
0 & 1 & 0 \\
0 & 0 & -1
\end{bmatrix} \;, \\
\label{matZ}
Z=
\begin{bmatrix}
-1 & 0 & 0 \\
0 & -1 & 0 \\
0 & 0 & 1
\end{bmatrix} \;,
\end{eqnarray}
useful for the derivations of Section~\ref{qbswitch_sec}.


\end{document}